\documentclass[%
twocolumn,
superscriptaddress,
 amsmath,amssymb,
 aps,
pra,
]{revtex4-2}

\usepackage{graphicx}
\usepackage{dcolumn}
\usepackage{bm}
\usepackage{color}
\usepackage{ulem}
\usepackage{stackengine}
\usepackage{braket}
\usepackage{multirow}

\newcommand{\figref}[1]{Fig.~\ref{fig:#1}}
\newcommand{\secref}[1]{Sec.~\ref{sec:#1}}
\newcommand{\equref}[1]{Eq.~(\ref{eq:#1})}
\newcommand{\tabref}[1]{Table~\ref{tab:#1}}

\begin{document}

\title{
Universality of Efimov states in highly mass-imbalanced cold-atom mixtures with van der Waals and dipole interactions
}%
\author{Kazuki Oi}
\affiliation{Department of Physics, Tohoku University, Sendai 980-8578, Japan}
\affiliation{Department of Engineering Science, University of Electro-Communications, Chofu, Tokyo 182-8585, Japan}
\author{Pascal Naidon}
\affiliation{Few-Body Systems Physics Laboratory, RIKEN Nishina Centre, RIKEN, Wako, 351-0198 Japan }
\author{Shimpei Endo}
\email{shimpei.endo@uec.ac.jp}
\affiliation{Department of Engineering Science, University of Electro-Communications, Chofu, Tokyo 182-8585, Japan}

\date{\today}


\begin{abstract}
We study three-body systems in a mass-imbalanced two-component cold-atom mixture, and we investigate the three-body parameter of their Efimov states for both bosonic and fermionic systems, with a major focus on the Er-Er-Li Efimov states. For a system interacting solely via van der Waals interactions, the van der Waals universality of the three-body parameter is analytically derived using the quantum defect theory. 
With the addition of a perturbative dipole interaction between the heavy atoms, the three-body parameters of the  bosonic and fermionic Efimov states are found to behave differently. When the dipole interaction is as strong as the van der Waals interaction, corresponding to realistic Er-Er-Li Efimov states, we show that the van der Waals universality persists once the effects of the non-perturbative dipole interaction are {\it renormalized} into the $s$-wave and $p$-wave scattering parameters between the heavy atoms.
For a dipole interaction much stronger than the van der Waals interaction, we find that the universality of the Efimov states can be alternatively characterized by a quasi-one-dimensional scattering parameter due to a strong anisotropic deformation of the Efimov wave functions. Our work thus clarifies the interplay of isotropic and anisotropic forces in the universality of the Efimov states. Based on the renormalized van der Waals universality, the three-body parameter is estimated for specific isotopes of Er-Li cold-atom mixtures.

\end{abstract}

\maketitle



\section{Introduction}
Recent progress in the field of cold atoms has significantly advanced our understanding of strongly correlated quantum systems and their universal behaviors. A prime example of quantum phenomenon realized in cold atoms is the Efimov effect~\cite{efimov1970energy,efimov1973energy,naidon2017efimov,RevModPhys.89.035006,d2018few,Braaten2006259}, which allows the existence of weakly bound three-body states featuring a discrete scale invariance. While these Efimov states have been theoretically suggested to appear universally for various physical systems~\cite{naidon2017efimov,AnnRev_HamPlatt,nishida2013efimov}, they require the particles to interact with large $s$-wave scattering lengths. This condition can be achieved in cold atoms by using a Feshbach resonance~\cite{inouye1998observation,chin2010feshbach}, which enables the precise control of the $s$-wave scattering length between the atoms. Efimov states have been observed in systems of identical bosons~\cite{kraemer2006evidence}, mass-imbalanced bosonic mixtures~\cite{PhysRevLett.112.250404,PhysRevLett.113.240402,PhysRevLett.115.043201}, and three-component Fermi systems~\cite{PhysRevLett.101.203202,PhysRevLett.103.130404,PhysRevLett.105.023201}. An Efimov state has also been observed as the excited state of the $^4$He trimer~\cite{kunitski2015observation}, due to a naturally large $s$-wave scattering length between $^4$He atoms. Importantly, the Efimov states observed for various different atomic species~\cite{PhysRevLett.107.120401} and internal spin states~\cite{gross2011study} have been found to have almost the same physical property: the three-body parameter, characterizing the binding energy and hence the size of the Efimov states, has been found to scale universally with the van der Waals length between the atoms. This universality of the three-body parameter has been termed ``van der Waals universality'' and actively investigated both theoretically and experimentally for identical bosons~\cite{PhysRevLett.107.120401,gross2011study,pascaleno3BP1,pascaleno3BP2,PhysRevLett.108.263001} as well as mass-imbalanced bosonic mixtures~\cite{PhysRevLett.109.243201,PhysRevA.95.062708}; It has been demonstrated to hold true except for systems close to a narrow Feshbach resonance~\cite{PhysRevLett.111.053202,PhysRevLett.123.233402,PhysRevLett.125.243401,johansen2017testing,schmidt2012efimov,PhysRevA.103.052805,PhysRevLett.93.143201,PhysRevLett.100.140404,PhysRevA.86.052516}.

Efimov physics has also been explored in fermionic systems. The repulsion between identical fermions due to the Pauli exclusion 
competes with the Efimov attraction, and leads to a richer few-body physics.
Theoretically, Efimov trimers have been predicted to appear in a mass-imbalanced two-component fermionic mixture when the mass ratio is large: $M/m>13.6\ldots$~\cite{efimov1973energy,PhysRevA.67.010703}. With such a large mass imbalance, the Efimov attraction surpasses the repulsion and forms the Efimov trimers, in stark contrast to a moderate mass ratio $8.1\ldots<M/m<13.6\ldots$ for which there exist ``Kartavtsev-Malykh trimers''~\cite{kartavtsev2007low,PhysRevLett.103.153202} and their crossover into the Efimov trimers~\cite{PhysRevA.86.062703}, and a small mass ratio $M/m<8.1\ldots$ for which no trimer appears.  While these fermionic trimers are formed in the $L^\Pi=1^-$ state, distinct from the $L=0^+$ state of identical bosons, they require a fermionic cold atom mixture with a well-controlled Feshbach resonance. Until recently, the $^{40}$K-$^6$Li mixture has been the major mass-imbalanced Fermi system. The mass ratio, however, is smaller than that required for the appearance of the trimer, and only a precursor of the trimer formation has been observed~\cite{PhysRevLett.112.075302}. Furthermore, the Feshbach resonance of $^{40}$K-$^6$Li is of a relatively narrow character~\cite{naik2011feshbach,chin2010feshbach}. Recently, Feshbach resonances have been realized in highly mass-imbalanced Fermi mixtures of Er-Li~\cite{ErLiFR1,ErLiFR2} and Cr-Li~\cite{CrLiFR}, where multiple broad resonances have been found. This opens the possibility of observing the trimer physics of fermions in cold atoms. In particular, Er-Li is well inside the Efimov regime, and it is expected to be the first system to exhibit a ``rotating'' Efimov state, i.e., Efimov state with non-zero orbital angular momentum $L=1$. One major challenge of the Er-Li mixture is that there is a strong dipole-dipole interaction between the Er atoms~\cite{PhysRevLett.108.210401,PhysRevLett.112.010404}.
The dipole interaction couples states with different angular momentum, so that the orbital angular momentum $L$ is no longer a good quantum number~\cite{PhysRevLett.106.233201,PhysRevLett.107.233201}. Since the dipole interaction is as strong as the van der Waals interaction, we are faced with the question of how the interplay of dipole and van der Waals interaction can modify the universal properties of the Efimov states: ``Is the three-body parameter in the Er-Li system universal?''

We address this question by investigating the Efimov states in a highly mass-imbalanced three-body system. Using the Born-Oppenheimer approximation valid for a highly mass-imbalanced system, we study the universality of the three-body parameter,  i.e., the binding energy of the Efimov states in the unitary limit. To elucidate the role of the statistics of the heavy atoms, we study both bosonic and fermionic systems with a variable strength of dipole interaction. First, we study the three-body system interacting solely with a van der Waals interaction. Using the quantum defect theory (QDT), the three-body parameter is explicitly expressed as a universal function of the van der Waals length and $s$-wave scattering length ($p$-wave scattering volume) between the heavy atoms for bosons (fermions), respectively. 
Second, for a system interacting with the van der Waals interaction and a weak dipole interaction, our perturbative analysis, together with numerical calculations, show that the bosonic and fermonic Efimov states are affected in a distinct manner by the dipole interaction. Third, for moderate dipole interaction strengths, including the realistic dipole strengths between Er atoms, our coupled-channel numerical calculations vindicate the universality of the three-body parameter. For the bosonic system, the van der Waals universality formula is found to persist even though the dipole interaction is so large that the $s$-wave scattering length and the binding energy are significantly modified; once the effects of the dipole interaction are {\it renormalized} into the $s$-wave scattering length, the QDT formula derived for a purely van der Waals system holds true, so that the $s$-wave scattering length between the heavy atoms universally characterizes the three-body parameter. Based on this renormalized van der Waals universality, we estimate the three-body parameters of the Efimov states for some specific isotopes of Er-Li systems. For the fermionic systems, the $p$-wave scattering volume is no longer a viable physical parameter, and we find an alternative low-energy scattering parameter which characterizes the universality of the three-body parameter. Finally, for a dipole interaction much stronger than the van der Waals interaction, the above universal description with the three-dimensional scattering parameters should break down, but we find that quasi-one-dimensional scattering parameters between the heavy atoms can be used to universally describe the Efimov states in both bosonic and fermionic systems.

This paper is organized as follows: in \secref{BO}, after introducing the Born-Oppenheimer description of the mass-imbalanced three-body system, our QDT analysis of non-dipolar system is presented. In \secref{Results}, we present the results of our numerical calculations and compare them with our analytical results for non-dipolar, weak dipole, and moderate dipole interactions, followed by the analysis in the limit of strong dipole interactions. In \secref{TBP_ErLi}, we present our estimates of the three-body parameters of some Er-Li isotopes. We conclude and present future perspectives in \secref{Concl}. Throughout the paper, the natural unit $\hbar=1$ is used.


\section{\label{sec:BO}Born-Oppenheimer description of highly mass-imbalanced 3-body system}
\subsection{General Formulation}
We consider a three-body system of two heavy particles (mass $M$) which interact resonantly with a light particle (mass $m$). The heavy particles are assumed to be either identical spin-polarized bosons or fermions, to model $^{166,168,170}$Er and $^{167}$Er atoms, respectively. As the light particle and heavy particles interact with a large $s$-wave scattering length $a_{\mathrm{HL}}$, a condition for the Efimov states to appear, we model the light-heavy interaction by the zero-range contact interaction~\cite{bethe1935scattering}. On the other hand, the interaction between the heavy particles is assumed to be the sum of van der Waals and dipole interactions. This amounts to neglecting the van der Waals interaction length scale between the light and heavy particles, which should be a reasonable assumption because the van der Waals length between the heavy atoms is larger than that between the light-heavy atoms. Throughout this paper, we focus on the unitary limit $1/a_{\mathrm{HL}} =0$, and model the light-heavy interaction via the unitary Bethe-Peierls boundary condition~\cite{bethe1935scattering}. We note that this treatment of the inter-species interaction is valid in describing broad Feshbach resonances, but less so for narrower ones~\cite{chin2010feshbach}

In a highly mass-imbalanced system, the Born-Oppenheimer approximation provides a good description of the Efimov physics. Indeed, it successfully reproduces the Efimov scale factor, and the critical mass ratio for the Efimov effect $M/m=13.99\ldots$~\cite{efimov1973energy} in good agreement with the exact value $M/m=13.60\ldots$~\cite{PhysRevA.67.010703}. The Born-Oppenheimer approximation has also been confirmed to reproduce well the van der Waals universality of the Efimov states without the dipole interaction for mass ratios $14 \lesssim M/m \lesssim 29$~\cite{PhysRevLett.109.243201}. It is therefore a good approximation for an Er-Li mixture, whose mass ratio is $M/m\approx 28$. The light particle's Schr\"{o}dinger equation is solved analytically and leads to the induced attractive interaction between the heavy particles $-\Omega^2/2mr^2$, where $\Omega=0.5671\ldots$~\cite{naidon2017efimov}. The Schr\"{o}dinger equation between the heavy particles then reads
\begin{equation}
 \label{eq:BOeq}  \left[-\frac{\nabla_{\bm{r}}^2}{M} -\frac{\Omega^2}{2mr^2}-\frac{C_6}{r^6}+\frac{C_{dd}(1-3\cos^2\theta)}{r^3}\right]\psi(\bm{r})=E\psi(\bm{r}),
\end{equation}
where $\theta$ is an angle measured from the $z$ axis taken to be parallel to the orientation of the dipoles (parallel to the external magnetic field). Here, $C_6$ and $C_{dd}$ are the van der Waals and dipole coefficients, from which we can define the van der Waals length $r_{\mathrm{vdw}}= \dfrac{1}{2}(MC_6)^{1/4}$ and the dipole length $a_{dd}=MC_{dd}/3 $ between the heavy atoms.

The binding energy of the Efimov states (i.e., three-body parameter) can be obtained by solving \equref{BOeq}. It needs to be supplemented with a short-range boundary condition. For this purpose, we use a hard wall condition at $R_{\mathrm{min}}$: $\psi(r=R_{\mathrm{min}})=0$. We will see later that the results are insensitive to the way the boundary condition is introduced as long as $R_{\mathrm{min}}$ is small enough. As the dipole interaction couples different partial waves, the angular momentum is no longer a good quantum number, but the parity $\Pi$ and the $z$ component of the angular momentum $M_z$ are good quantum numbers. While various values of $M_z$ are allowed in general, we restrict ourselves to the $M_z=0$ channel in this paper. For the bosonic heavy atoms, the $M_z^\Pi=0^+$ state is allowed by the symmetry. This state involves the $L=0,2,4,...$ angular momentum channels, among which the $L^\Pi=0^+$ channel shows the Efimov effect and the others do not for the Er-Li mass ratio~\cite{efimov1973energy,kartavtsev2007low,endo2011universal,Helfrich_2011,naidon2017efimov}. For the fermionic heavy atoms, the $M_z^\Pi=0^-$ channel is allowed by the symmetry. This state involves $L=1,3,5,...$ angular momentum channels, among which the $L^\Pi=1^-$ channel shows the Efimov effect and the others do not for the mass ratio in question~\cite{efimov1973energy,kartavtsev2007low,endo2011universal,Helfrich_2011,naidon2017efimov}. Therefore, in the following, we study the $M_z^\Pi=0^+$ and  $M_z^\Pi=0^-$ states for the bosonic and fermionic Er atoms, respectively.



\subsection{\label{sec:qdt_vdw}Quantum Defect Theory for non-dipolar system}

For typical cold atomic species, $r_{\mathrm{vdw}} \gg a_{dd}$ so that the dipole interaction is negligible in \equref{BOeq}. While this is not the case for Er atoms, it still provides us with a basis for understanding the universal behavior of the Efimov states with the dipole interactions. Therefore, in this section we analyze 
\begin{equation}
 \label{eq:BOeq_vdwonly}  \left[-\frac{\nabla_{\bm{r}}^2}{M}-\frac{\Omega^2}{2mr^2}-\frac{C_6}{r^6}\right]\psi(\bm{r})=E\psi(\bm{r})
\end{equation}
with a short-range boundary condition $\psi(|\bm{r}|=R_{\mathrm{min}})=0$. While this equation was studied numerically in Ref.~\cite{PhysRevA.95.062708} and the van der Waals universality of the Efimov states was clarified for the mass-imbalanced system, we show below that \equref{BOeq_vdwonly} can be solved analytically and provide explicit formulas for the binding energies and wave functions of the Efimov states.

Without the dipole interaction, the angular momentum is a good quantum number, and \equref{BOeq_vdwonly} is essentially a single-channel Schr\"{o}dinger equation with $1/r^{6}$ potential:
\begin{equation}
 \label{eq:SingleChanBOeq_vdwonly}  \left[-\frac{1}{M} \frac{d^2}{dr^2}+ \frac{s_\ell^2-\frac{1}{4}}{Mr^2}-\frac{C_6}{r^6}\right]u_\ell(r)=Eu_\ell(r)
\end{equation}
with  $\psi(\bm{r})=\dfrac{u_\ell(r)}{r} Y_{\ell m}(\bm{r})$ and 
\begin{equation}
s_\ell^2 =  \left(\ell+\frac{1}{2}\right)^2-\frac{M}{2m}\Omega^2 .
\end{equation}
Since we are interested in the Efimov states, we focus on the $\ell=0$ ($\ell=1$) channel where $s_\ell^2 <0$ for the mass ratio of interest $M/m \approx 28$ for bosons (fermions), respectively~\cite{efimov1973energy,kartavtsev2007low,naidon2017efimov,endo2011universal,Helfrich_2011}. Equation~(\ref{eq:SingleChanBOeq_vdwonly}) can be solved analytically using Ref.~\cite{PhysRevA.58.1728}, with a remark that the centrifugal term is attractive due to the Efimov attraction. In other words, the angular momentum in Ref.~\cite{PhysRevA.58.1728} should be generalized to an imaginary number. We can then apply the quantum defect theory (QDT)~\cite{PhysRevA.64.010701,BoGao2004} to obtain the analytical behavior of the Efimov states; using the QDT parameter $K^c$ introduced as
\begin{equation}
\label{eq:QDT_uKc_definition}u_\ell(r ) = f_{\ell}^c (r) - K^c g^c_{\ell}(r) ,
\end{equation}
where $f_{\ell}^c $ and $g_{\ell}^c $ are two independent solutions of \equref{SingleChanBOeq_vdwonly}. The binding energies of the Efimov states can be obtained by imposing $u_\ell(r\rightarrow \infty)=0$, from which we find~\cite{PhysRevA.64.010701} 
\begin{equation}
   \label{eq:chi_func}
   K^c=\frac{W^c_{f-}}{W^c_{g-}}=\frac{(1+M_{\ell})\sin\dfrac{\pi\nu}{2}\cos\theta_{\ell}+(1-M_{\ell})\cos\dfrac{\pi\nu}{2}\sin\theta_\ell}{(1-M_{\ell})\cos\dfrac{\pi\nu}{2}\cos\theta_{\ell}-(1+M_{\ell})\sin\dfrac{\pi\nu}{2}\sin\theta_\ell} .
\end{equation}
The parameters $\theta_\ell$, $\nu$, and $M_{\ell}$ can be obtained from the standard procedure of solving the van der Waals two-body problem (see Appendix.~\ref{app:vdw2body}). At low energy $|E| \ll 1/Mr^2_{\mathrm{vdw}}$, by performing the low-energy expansion in a similar manner to Refs.~\cite{BoGao2004,PhysRevA.80.012702}, we find (see Appendix.~\ref{app:QDTexpansion} )
 \begin{equation}
\label{eq:vddUniversalEfimovBEformula}  |E|= \frac{4}{Mr_{\rm vdw}^2}\exp\left[\frac{2}{|s_\ell|}\left\{\arctan\left(\frac{K^c}{\tanh\dfrac{\pi|s_\ell|}{4}}\right)+\xi_\ell \right\}\right]e^{-\frac{2n\pi}{|s_\ell|}} ,
\end{equation}
where $\xi_\ell = \arg\left[\Gamma\left(\frac{i|s_\ell|}{2}\right)\Gamma(1+i|s_\ell|)\right]$ and $n$ is an arbitrary integer satisfying $|E| \ll 1/Mr^2_{\mathrm{vdw}}$.

The QDT parameter $K^c $ is formally a matrix acting on multiple channels, but it simplifies as a scalar parameter for the single-channel scattering. Its value is related to the short-range phase, hence the short-range boundary condition. Indeed, at short distance, one can substitute the low-energy form of the wave function \equref{vdw_only_0_energy_sol} because the energy is much smaller than the potential, and imposing $u_\ell(r =R_{\mathrm{min}})=0$, one finds
\begin{align}
K^c &= - \tanh\left(\frac{\pi|s_\ell|}{4}\right)\frac{\mathrm{Re}\left[J_{\frac{s_\ell}{2}}\left(\Phi \right)\right]}{\mathrm{Im}\left[J_{\frac{s_\ell}{2}}\left(\Phi \right)\right]}\label{eq:KcRminrelation1}\\
& \simeq  - \frac{1}{\tan\left[\Phi - \frac{\pi}{4} \right]}\label{eq:KcRminrelation2},
\end{align}
where $\Phi  = 2r_{\rm vdw}^2 /R^2_{\mathrm{min}}$. In the second line, we have used the asymptotic form of the Bessel function (or equivalently \equref{fc_gc_shortrange}) valid for $R_{\mathrm{min}} \ll r_{\rm vdw}$. This relation between $R_{\mathrm{min}}$ and $K^c$, substituted into \equref{vddUniversalEfimovBEformula} gives an explicit analytical formula of the Efimov binding energies as a function of $R_{\mathrm{min}}$. The wave function of the Efimov states is also analytically obtained as ($x  \equiv2\left(r_{\rm vdw}/r\right)^2$)
\begin{align}
u_\ell(r ) &=  \sqrt{ \frac{r}{2} }\left( \frac{ \mathrm{Re}\left[J_{\frac{s_\ell}{2}}\left(x \right)\right] }{ \cosh \frac{\pi |s_\ell|}{4}} + K^c \frac{\mathrm{Im}\left[J_{\frac{s_\ell}{2}}\left(x \right)\right] }{ \sinh \frac{\pi |s_\ell|}{4}}   \right)\\
& = \sqrt{ \frac{r}{2}}\frac{\mathrm{Re}\left[J_{\frac{s_\ell}{2}}\left(\Phi  \right)\right]}{ \cosh \frac{\pi |s_\ell|}{4}} \left( \frac{\mathrm{Re}\left[J_{\frac{s_\ell}{2}}\left(x \right)\right]}{\mathrm{Re}\left[J_{\frac{s_\ell}{2}}\left(\Phi \right)\right]} -\frac{ \mathrm{Im}\left[J_{\frac{s_\ell}{2}}\left(x \right)\right] }{\mathrm{Im}\left[J_{\frac{s_\ell}{2}}\left(\Phi  \right)\right]}  \right),
\end{align}
where the log-periodic behavior characteristic of the Efimov states appears from the oscillation of the imaginary-index Bessel function.

While the $R_{\mathrm{min}}$ dependence of the binding energy and wave function may look like a non-universal short-range dependence of the Efimov states, we can recast it into universal forms by using the relation between $R_{\mathrm{min}}$ and the two-body universal scattering parameter between the heavy atoms; if we consider the two-body problem of the two heavy atoms
\begin{equation}
\label{eq:vdw2body_scheq}  \left[-\frac{1}{M} \frac{d^2}{dr^2} + \frac{\ell (\ell+1)}{Mr^2}-\frac{C_6}{r^6}\right]u_\ell(r)=Eu_\ell(r) , 
\end{equation}
the only difference with \equref{BOeq_vdwonly} is the presence/absence of the $1/r^2$ attraction. At short distance $r \sim R_{\mathrm{min}} \ll r_{\mathrm{vdw}}$, this attraction is much smaller than the van der Waals force so that the two equations are essentially the same. More physically, the short-range phase of the three-body system is dominated by the potential between the heavy atoms. This assumption should be particularly true for a highly mass-imbalanced system, because due to larger atomic polarizability of the heavy atoms, their van der Waals force and higher-order forces are much stronger than those for the light atom. Therefore, the QDT parameter of the three-body system can be approximated by that of two heavy atoms $K^c_{\mathrm{3body}}\simeq K^c_{\mathrm{2body}}$, which amounts to adopting the same short-range cutoff $R_{\mathrm{min}}$ value for the two- and three-body systems. While this assumption is plausible for our model, it can be compromised by finite-range effects of the heavy-light atoms' interaction. It may also be invalidated by the non-adiabatic effects beyond the Born-Oppenheimer approximation, which break down the single-channel description.

The van der Waals two-body problem \equref{vdw2body_scheq} with the QDT condition \equref{QDT_uKc_definition} has been solved in Refs.~\cite{BoGao2004,PhysRevA.80.012702}, and the following relations between the the QDT parameter and the two-body scattering parameters have been obtained:
\begin{align}
   \label{eq:aHH_only_vdw}
   \frac{a^{(\mathrm{HH})}}{r_{\rm vdw}}& =\frac{4\pi}{\Gamma^2\left(\frac{1}{4}\right)}\left[1+\frac{1+K^c\tan\dfrac{\pi}{8}}{K^c-\tan\dfrac{\pi}{8}}\right] \\
   \label{eq:vp_only_vdw}
  \frac{v_p^{(\mathrm{HH})}}{r^3_{\rm vdw}} & =-\frac{1}{3}\frac{\Gamma\left(\frac{1}{4}\right)}{\Gamma\left(\frac{7}{4}\right)}\frac{1+\tan\dfrac{3\pi}{8}-\left(1-\tan\dfrac{3\pi}{8}\right)K^c}{\sqrt{2}\left(\tan\dfrac{3\pi}{8}-K^c\right)},
\end{align}
where $ a^{(\mathrm{HH})}$ and $v_p^{(\mathrm{HH})}$ are the $s$-wave scattering length and $p$-wave scattering volume between the heavy atoms, respectively. Noting that the bosonic (fermionic) heavy atoms interact dominantly with the $s$-wave scattering length ($p$-wave scattering volume) at low energy, and substituting Eqs.~(\ref{eq:aHH_only_vdw}) and~(\ref{eq:vp_only_vdw}) into \equref{vddUniversalEfimovBEformula}, the binding energies of the Efimov states are represented as
\begin{widetext}
\begin{equation}
  \label{eq:vdw_efimov_aHH_and_binding_energy}
   |E|=\frac{4}{Mr_{\rm vdw}^2}\exp\left[\frac{2}{|s_0|}\left\{\arctan\left(\frac{1}{\tanh\dfrac{\pi|s_0|}{4}}\frac{\dfrac{a^{(\mathrm{HH})}}{r_{\rm vdw}}\tan\dfrac{\pi}{8}+\dfrac{4\pi}{\Gamma^2(\frac{1}{4})}\left(1-\tan\dfrac{\pi}{8}\right)}{\dfrac{a^{(\mathrm{HH})}}{r_{\rm vdw}}-\dfrac{4\pi}{\Gamma^2(\frac{1}{4})}\left(1+\tan\dfrac{\pi}{8}\right)}\right)+\xi_0 \right\}\right]e^{-\frac{2n\pi}{|s_0|}}
\end{equation}
for the bosonic heavy atoms ($\ell=0$), and 
\begin{equation}
   \label{eq:vdw_efimov_vHH_and_binding_energy}
   |E|=\frac{4}{Mr_{\rm vdw}^2}\exp\left[\frac{2}{|s_1|}\left\{\arctan\left(\frac{1}{\tanh\dfrac{\pi|s_1|}{4}}\frac{\dfrac{v_p^{(\mathrm{HH})}}{r_{\rm vdw}^3}\tan\dfrac{3}{8}\pi+\dfrac{1}{3\sqrt{2}}\dfrac{\Gamma(\frac{1}{4})}{\Gamma(\frac{7}{4})}\left(1+\tan\dfrac{3}{8}\pi\right)}{\dfrac{v_p^{(\mathrm{HH})}}{r_{\rm vdw}^3}+\dfrac{1}{3\sqrt{2}}\dfrac{\Gamma(\frac{1}{4})}{\Gamma(\frac{7}{4})}\left(1-\tan\dfrac{3}{8}\pi\right)}\right)+\xi_1 \right\}\right]e^{-\frac{2n\pi}{|s_1|}}
\end{equation}
for the fermionic heavy atoms ($\ell=1$), where $n$ is an arbitrary integer such that $|E|\ll 1/Mr_{\rm vdw}^2$.
\end{widetext}

As we will see in \secref{res_vdw}, these analytical formulas excellently reproduce the numerical results for non-dipolar systems. These analytical formulas suggest that the binding energy of the Efimov states in the unitary limit, i.e. the three-body parameter, is insensitive to the short-range details of the system but universally determined by the van der Waals length and the $s$-wave scattering length ($p$-wave scattering volume) for the bosonic (fermionic) Efimov states. This van der Waals universality has been experimentally demonstrated for equal-mass~\cite{PhysRevLett.107.120401,gross2011study} and mass-imbalanced bosonic systems~\cite{johansen2017testing} close to a broad Feshbach resonance.  The van der Waals universality has also been demonstrated theoretically with numerical calculations for  equal-mass~\cite{PhysRevLett.108.263001,pascaleno3BP1,pascaleno3BP2} and mass-imbalanced bosonic systems~\cite{PhysRevLett.109.243201,PhysRevA.95.062708}. In particular, the same Born-Oppenheimer equation~(\ref{eq:BOeq_vdwonly}) with the same short-range boundary condition has been employed in Ref.~\cite{PhysRevA.95.062708} and solved numerically to study the universality of the Efimov states. Our QDT analysis provides an analytical demonstration of the van der Waals universality of hetero-nuclear Efimov states; namely, the assumption of $K^c_{\mathrm{3body}}\simeq K^c_{\mathrm{2body}}$ is pivotal in universally characterizing the three-body parameter with the two-body scattering parameters between the heavy atoms. As we use the single-channel QDT, Eqs.~(\ref{eq:vdw_efimov_aHH_and_binding_energy}) and~(\ref{eq:vdw_efimov_vHH_and_binding_energy}) are invalid for describing the Efimov states near a narrow Feshbach resonance, which have been demonstrated both experimentally~\cite{PhysRevLett.111.053202,PhysRevLett.123.233402,PhysRevLett.125.243401,johansen2017testing} and theoretically~\cite{schmidt2012efimov,PhysRevA.103.052805,PhysRevLett.93.143201,PhysRevLett.100.140404,PhysRevA.86.052516} to deviate from the van der Waals universality; the multi-channel QDT~\cite{PhysRevA.72.042719,PhysRevA.84.042703} is necessary for capturing their sensitivity to the channel coupling and nature of the internal spin states.





\begin{figure}[!t]
	\centering
	\includegraphics[width=0.9\linewidth]{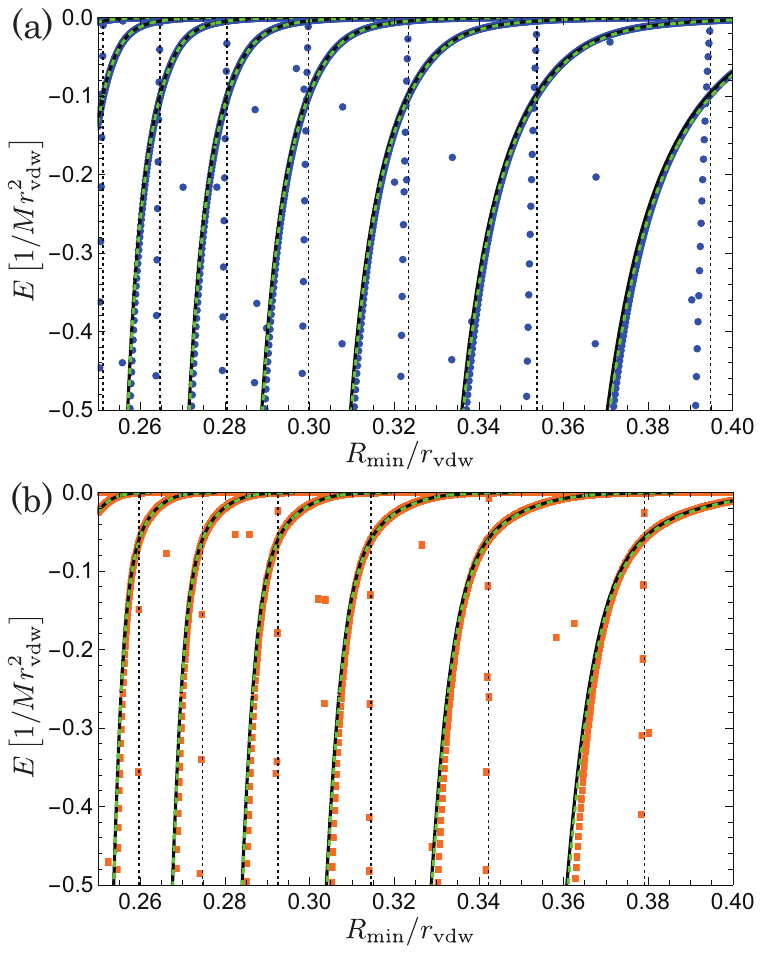} 
\caption{Binding energy (in the van der Waals energy unit $1/M r_{\mathrm{vdw}}^2$) of the Efimov states with the unitary heavy-light interaction for (a) bosonic and (b) fermionic heavy atoms interacting via the van der Waals interaction. The mass ratio are (a)  $M/m=27.5855...$ and (b) $M/m=27.7520...$, corresponding to $^{166}$Er-$^6$Li and $^{167}$Er-$^6$Li, respectively. The numerical results  for the bosons (blue circles) and fermions (orange squares) with $\ell_{\mathrm{max}}=20$ are shown, together with the analytical curves of \equref{vddUniversalEfimovBEformula} with \equref{KcRminrelation1} (black solid) and \equref{KcRminrelation2} (green dashed). The vertical thin dotted lines (black) denote the thresholds at which the higher angular momentum bound states of $\ell=2$ ($\ell=3$) start to appear for the bosons (fermions), obtained from Eqs.~(\ref{eq:threshold_higherZeroE}) and~(\ref{eq:KcRminrelation1}).}
\label{fig:VdWOnlyERmin}
\end{figure}

\section{\label{sec:Results}Numerical Results}
We show the numerical solutions of \equref{BOeq} in this section. As we are particularly interested in Er-Er-Li three-body systems, we perform the calculations with Er-Li mass ratio: $M/m=27.5855...$ for the bosonic system corresponding to $^{166}$Er-$^6$Li, and $M/m=27.7520...$ for the fermions corresponding to $^{167}$Er-$^6$Li, respectively. The differential equation~(\ref{eq:BOeq}) is discretized in the log-scaled coordinate space between $R_{\mathrm{min}}$ and $R_{\mathrm{max}}$, where $R_{\mathrm{max}}$ is typically taken as $\sim$~400-1000~$r_{\mathrm{vdw}}$. We find that 4000 grid points and $\ell_{\mathrm{max}}=$10-20 are enough to attain convergence for most cases, while we take up to 30000 grid points and $\ell_{\mathrm{max}}\approx 40$ for some cases of stronger dipole interactions.

\subsection{\label{sec:res_vdw}Non-dipolar case $a_{dd}=0$}

\begin{figure}[!t]
	\centering
	\includegraphics[width=1.0\linewidth]{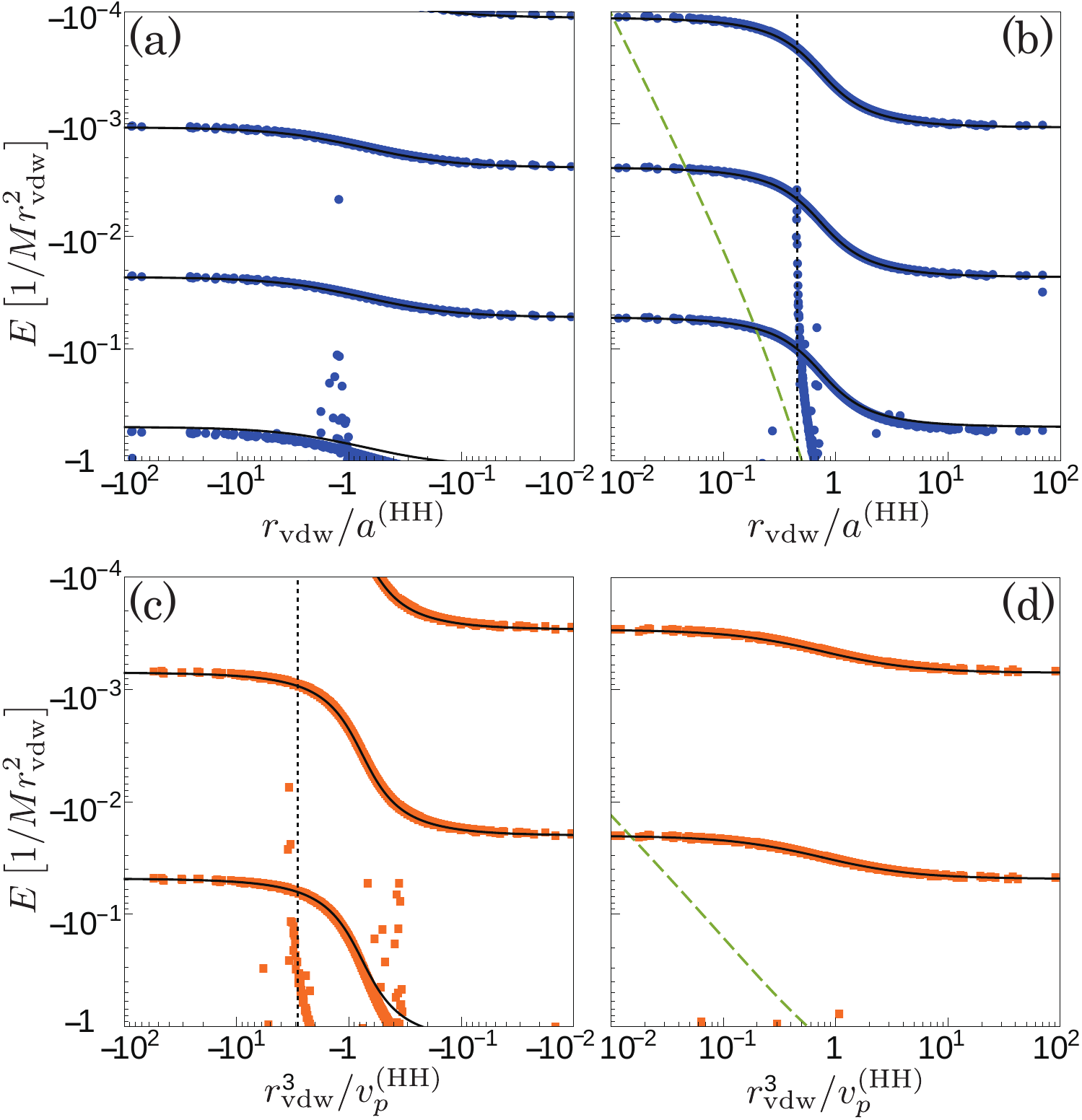} 
\caption{Binding energy of the Efimov states plotted against (a)(b) the $s$-wave scattering length $a^{(\mathrm{HH})}$ for the bosons, and (c)(d) $p$-wave scattering volume $v_p^{(\mathrm{HH})}$ for the fermions, obtained by converting $R_{\mathrm{min}}$ in \figref{VdWOnlyERmin} into $a^{(\mathrm{HH})}$ and $v_p^{(\mathrm{HH})}$ using Eqs.~(\ref{eq:KcRminrelation1})~(\ref{eq:aHH_only_vdw}), and (\ref{eq:vp_only_vdw}). The left (right) pannels are negative (positive) scattering length/volume sides, which should smoothly connect with each other in the unitary limit $1/a^{(\mathrm{HH})}=0$,  $1/v_p^{(\mathrm{HH})}=0$ corresponding to the border between the left and right panels. The solid curves (black) are the analytical results of Eqs.~(\ref{eq:vdw_efimov_aHH_and_binding_energy}) and (\ref{eq:vdw_efimov_vHH_and_binding_energy}), and the dashed curves (green) are the dimer binding energy of the heavy atoms. The vertical dotted lines (black) are the same as those in \figref{VdWOnlyERmin}.}
\label{fig:VdWOnlyEaHHvHH}
\end{figure}

First, we show the results without the dipole interaction: in \figref{VdWOnlyERmin}, we show the binding energy of the Efimov states (in the van der Waals energy unit $1/M r_{\mathrm{vdw}}^2$) as the short-range boundary $R_{\mathrm{min}}$ is varied. As $R_{\mathrm{min}}$ gets smaller, the short-range van der Waals attraction increases, leading to the successive appearance of bound states and increase of the binding energy. The numerical data points almost perfectly agree with the analytical curves obtained with the QDT in \equref{vddUniversalEfimovBEformula}, particularly at small energy. There are a few points which significantly deviate from the analytical curves. They are ascribed to higher angular momentum states of $\ell=2,4,...$ for the bosons and $\ell=3,5,...$ for the fermions, respectively. The thresholds at which the $\ell=2$ ($\ell=3$) bound states start to appear, denoted by the vertical dotted lines in \figref{VdWOnlyERmin}~(a) and \figref{VdWOnlyERmin}~(b), agree with most of the stray data points, supporting this interpretation. A few stray points which agree with neither are ascribed to higher angular momentum channels $\ell\ge 4$ ($\ell\ge 5$). This classification of the states has also been confirmed from the shape of the wave functions. Because the higher angular momentum states are localized to shorter distances than the $s$-wave ($p$-wave) state owing to the centrifugal barrier, they are more sensitive to $R_{\mathrm{min}}$, resulting in the steeper binding energy curves. The binding energy curves of different partial waves do not render an avoided crossing but simply cross because the angular momentum is a good quantum number in the absence of the dipole interaction.

In \figref{VdWOnlyEaHHvHH}, we show the binding energies plotted against the $s$-wave scattering length $a^{(\mathrm{HH})}$ and $p$-wave scattering volume $v^{(\mathrm{HH})}$ between the heavy atoms, obtained from \figref{VdWOnlyERmin} and Eqs.~(\ref{eq:KcRminrelation1})~(\ref{eq:aHH_only_vdw})~(\ref{eq:vp_only_vdw}). The numerical data with different values of $R_{\mathrm{min}}$ collapse into universal curves. Therefore, the three-body parameter of the Efimov states is universally characterized by ($a^{(\mathrm{HH})}$, $r_{\mathrm{vdw}}$) for the bosons and ($v_p^{(\mathrm{HH})}$, $r_{\mathrm{vdw}}$) for the fermions, respectively. In other words, they are insensitive to the short-range details, but are  universally characterized by the two-body scattering parameters which describe the long-range asymptotic behavior of the two heavy atoms. The numerical data in \figref{VdWOnlyEaHHvHH} excellently agree with the curves of Eqs.~(\ref{eq:vdw_efimov_aHH_and_binding_energy}) and (\ref{eq:vdw_efimov_vHH_and_binding_energy}), vindicating our QDT analytical formulas. Some points in \figref{VdWOnlyEaHHvHH} deviating from the curves are higher angular momentum states, whose energies change rapidly as $R_{\mathrm{min}}$ hence $a^{(\mathrm{HH})}$ and $v_p^{(\mathrm{HH})}$ are varied; although they significantly deviate from the analytical curves of Eqs.~(\ref{eq:vdw_efimov_aHH_and_binding_energy}) and (\ref{eq:vdw_efimov_vHH_and_binding_energy}), the data points with different $R_{\mathrm{min}}$ still tend to collapse into a single curve (\figref{VdWOnlyEaHHvHH}~(b)(c)) due to the angular-momentum-insensitive nature of the short-range phase originating from $R_{\mathrm{min}} \ll r_{\mathrm{vdw}}$~\cite{PhysRevA.64.010701}. The marginal disagreement of the large binding energy region $|E|\simeq 1/Mr_{\mathrm{vdw}}^2$ in \figref{VdWOnlyEaHHvHH}~(c) is ascribed to the breakdown of the low-energy condition used to derive Eqs.~(\ref{eq:vdw_efimov_aHH_and_binding_energy}) and (\ref{eq:vdw_efimov_vHH_and_binding_energy}).

We note that the universal behaviors of the bosonic system in \figref{VdWOnlyEaHHvHH}~(a)(b) were found numerically in Ref.~\cite{PhysRevA.95.062708}; we have found here that a similar universal behavior holds true for the fermions (\figref{VdWOnlyEaHHvHH}~(c)(d)), together with their analytical descriptions in Eqs.~(\ref{eq:vdw_efimov_aHH_and_binding_energy})~(\ref{eq:vdw_efimov_vHH_and_binding_energy}).

In \figref{VdWOnlyEaHHvHH}, we also show the dimer energy of two heavy atoms (green dashed) obtained by solving \equref{vdw2body_scheq}. Wherever the Efimov trimers lie above the dimer energy, they are in fact resonant trimers embedded in the dimer-atom continuum~\cite{PhysRevA.95.062708}. In the Born-Oppenheimer approximation, however, the trimers appear as bound states because the channel describing the dimer-atom continuum is neglected.

\begin{figure}[!t]
	\centering
	\includegraphics[width=0.9\linewidth]{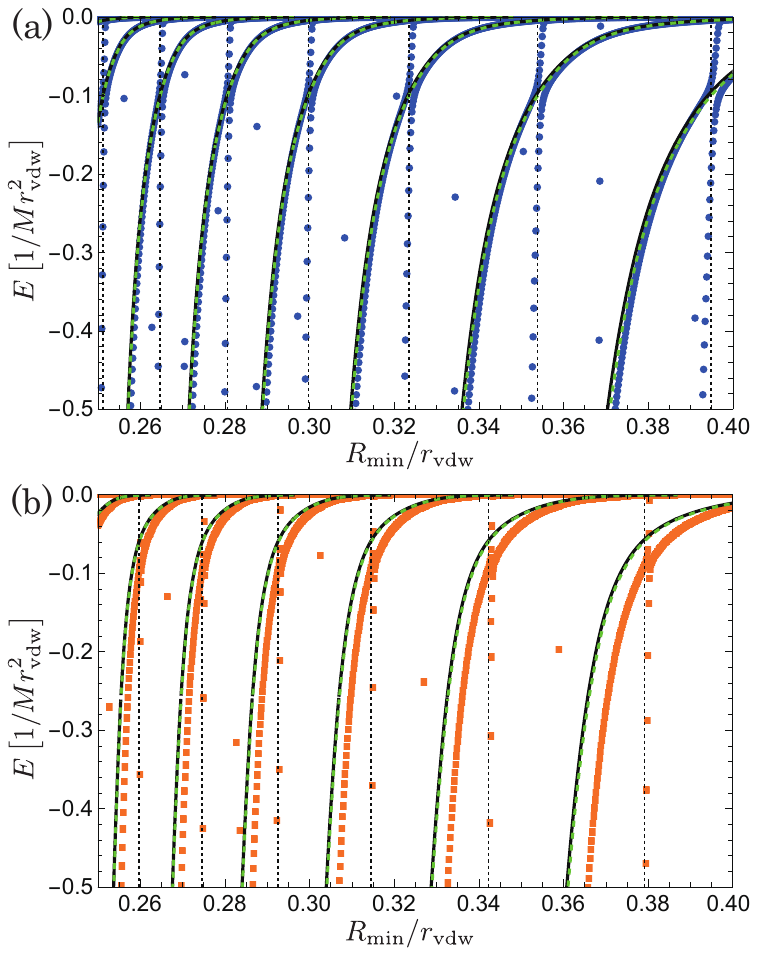} 
\caption{Binding energy of the Efimov states for (a) bosonic and (b) fermionic heavy atoms interacting via the van der Waals interaction and weak dipole interaction $a_{dd}/r_{\mathrm{vdw}}=0.40$. The other parameters and notations are the same as those in \figref{VdWOnlyERmin}.}
\label{fig:WeakDipoleERmin}
\end{figure}

\subsection{\label{sec:res_weak}Weak dipole interaction $a_{dd}\ll r_{\mathrm{vdw}}$}
We show in \figref{WeakDipoleERmin} the binding energies of the Efimov states in the presence of a weak dipole interaction. As the angular momentum is no longer a good quantum number, different partial wave states mix with each other, resulting in avoided crossings. We have confirmed that the shape of the wave function changes across the avoided crossing via a superposition of different partial waves. We have also confirmed that the width of the avoided crossings increases as $a_{dd}$ gets larger.

As shown in \figref{WeakDipoleERmin}, the numerical results generally get shifted toward larger binding energy. This can be understood by the perturbation theory: we have analytically found that the energy shift induced by the dipole interaction is $ E -E^{(0)}  \propto a_{dd}^2$ for the bosons predominantly in the $L=0$ channel, and $ E -E^{(0)}\propto a_{dd}$ for the fermions predominantly in the $L=1$ channel, with both negative shifts $ E -E^{(0)}<0$ (see Appendix.~\ref{app:perturbation}). The difference in the scaling originates from the absence (presence) of the diagonal matrix element of the dipole interaction for $L=0$ ($L=1$) states: $\langle L=0 |V_{dd}| L=0\rangle =0$, $\langle L=1 |V_{dd}| L=1\rangle \neq 0$. This is confirmed in \figref{pert_DeltaEaddplot}: the change in binding energy, i.e., three-body parameter of the Efimov states, scales quadratically for the bosons, and linearly for the fermions. Notably, in most previous studies on the Efimov effect, the role of the statistics of the particles has been rather marginal, at best to modify the value of $s_\ell$, and thereby change the absence/presence of the Efimov effect through the condition $s_\ell^2<0$ as the mass ratio is varied~\cite{efimov1973energy,PhysRevA.67.010703,endo2011universal,Helfrich_2011,naidon2017efimov,RevModPhys.89.035006,d2018few,Braaten2006259}. In other words, the statistics of the particles has been essentially irrelevant once its effect is incorporated into the value of $s_\ell$.  Figure~\ref{fig:pert_DeltaEaddplot} is one of the few examples where the quantum statistics of the particles plays an explicit role in qualitatively changing the universal behavior of the Efimov states.

\subsection{\label{sec:res_moderate}Moderate dipole interaction $a_{dd}\sim r_{\mathrm{vdw}}$}
\begin{figure}[!t]
	\centering
	\includegraphics[width=0.9\linewidth]{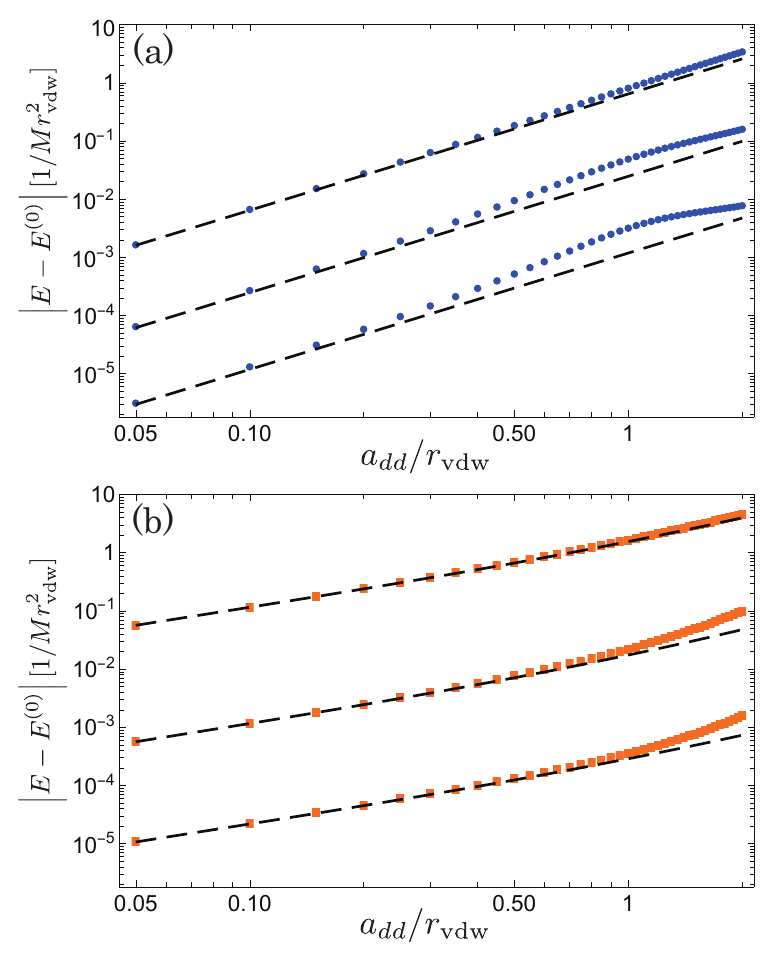} 
\caption{Energy shift induced via the dipole interaction with a fixed value of $R_{\mathrm{min}}/r_{\mathrm{vdw}}=0.40$ for (a) the bosons (blue circles) and (b) fermions (orange squares). The dashed lines (black) are the perturbation results in Eqs.~(\ref{eq:DeltaE_appEq_boson}) and (\ref{eq:DeltaE_appEq_fermion}), where the unperturbed energy $E^{(0)}$ and the matrix elements of \equref{app_def_udmatel} are evaluated with the energies and wave functions numerically obtained for the purely van der Waals system in \secref{res_vdw}.}
\label{fig:pert_DeltaEaddplot}
\end{figure}

Erbium atoms interact with a strong dipole interaction owing to their large magnetic moments: the dipole length of $^{166}$Er-$^{166}$Er is $a_{dd}=75.5a_0$, which is comparable to their van der Waals length $r_{\mathrm{vdw}}=65.5a_0$~\cite{PhysRevX.5.041029,Chomaz_2023} (see \tabref{result_list_TBP}). Figure~\ref{fig:pert_DeltaEaddplot} shows that this is away from the region where the perturbation theory works well $a_{dd}/r_{\mathrm{vdw}} \lesssim 0.5$.

\begin{figure*}[!t]
	\centering
	\includegraphics[width=1.0\linewidth]{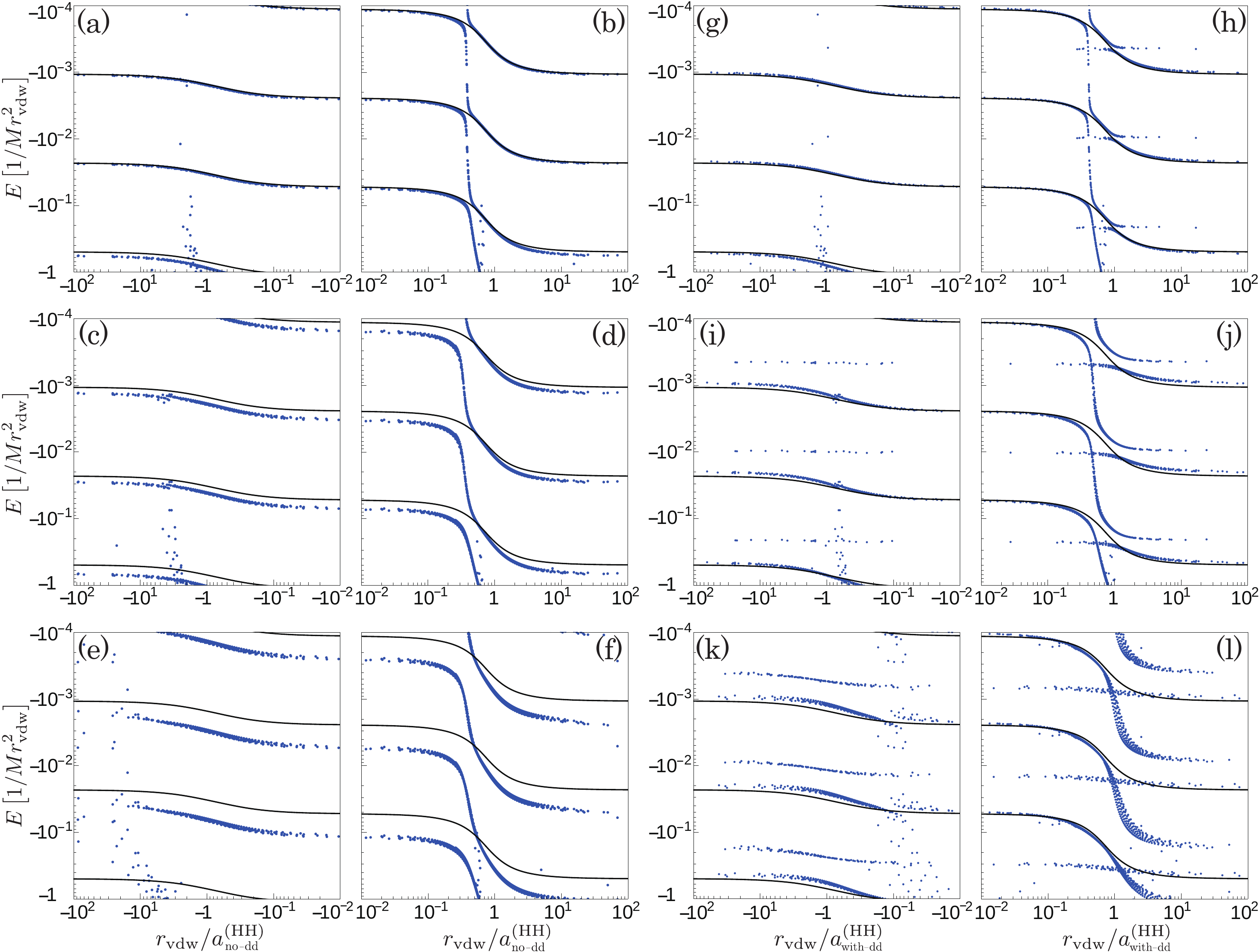} 
\caption{Binding energy of the bosonic Efimov states for the mass ratio  $M/m=27.5855...$ corresponding to $^{166}$Er-$^{166}$Er-$^6$Li system. Weaker dipole strength $a_{dd}/r_{\mathrm{vdw}}=0.40$ (top row (a)(b)(g)(h)), realistic dipole strength for $^{166}$Er-$^{166}$Er of $a_{dd}/r_{\mathrm{vdw}}=0.86755....$ (middle row (c)(d)(i)(j)), and stronger dipole strength $a_{dd}/r_{\mathrm{vdw}}=1.5$ (bottom row (e)(f)(k)(l)). (a)-(f) The horizontal axis is the $s$-wave scattering length between the heavy atoms estimated by \equref{KcRminrelation1}, i.e., without the dipole interaction. (g)-(l) The horizontal axis is the $s$-wave scattering length between the heavy atoms numerically obtained by solving the coupled-channel two-body equation in \equref{HeavyHeavy2bodyeq}. The left (right) panels in each column show the negative (positive) $s$-wave scattering length. The solid curve (black) is the analytical QDT formula for non-dipolar system in \equref{vdw_efimov_aHH_and_binding_energy}.
}
\label{fig:ModerateDipoleEaHHvHH_Boson}
\end{figure*}

We have therefore performed the numerical calculations for moderate strengths of the dipole interaction in \figref{ModerateDipoleEaHHvHH_Boson}(a)-(f); we first show the binding energy of the bosonic Efimov states plotted against the heavy-heavy $s$-wave scattering length estimated without the dipole interaction $a_{\mathrm{no-dd}}^{(\mathrm{HH})}$ (\equref{KcRminrelation1}). As the dipole strength is increased from a smaller ((a)(b)) to a larger ((e)(f)) value, the avoided crossing between the $\ell=0$ dominant state and the $\ell=2$ dominant state gets broader. Due to the admixture of the $\ell=0$ state, the slope of the $\ell=2$ dominant state appearing around $a^{(\mathrm{HH})}_{\mathrm{no-dd}}/r_{\mathrm{vdw}}\simeq 1$ gets gentler. Furthermore, the binding energies increase as the dipole interaction gets larger, which is consistent with the perturbative analysis in \secref{res_weak}. The deviation from the analytical curves of the non-dipolar system in \equref{vdw_efimov_aHH_and_binding_energy} (black curves) is so significant that they cannot well explain the numerical results of the realistic dipole interaction strength (middle row (c)(d)), let alone stronger ones (bottom row (e)(f)). Still, the binding energy data of different short-range parameters $R_{\mathrm{min}}$ collapse onto a single curve, suggesting the universality of the three-body parameter of the bosonic Efimov states even in the presence of non-perturbatively large dipole interactions.

\begin{figure*}[!t]
	\centering
	\includegraphics[width=1.0\linewidth]{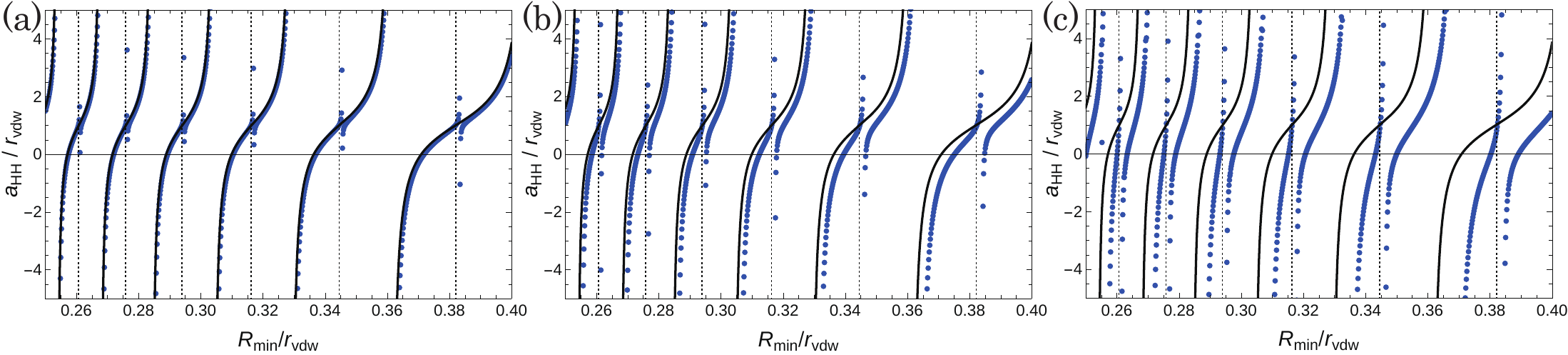} 
\caption{$S$-wave scattering length between the heavy atoms as $R_{\mathrm{min}}$ is varied, obtained numerically by solving \equref{HeavyHeavy2bodyeq}. The solid curve (black) is the $s$-wave scattering length without the dipole interaction $a^{(\mathrm{HH})}_{\mathrm{no-dd}}$, i.e., the analytical formula of Eqs.~(\ref{eq:aHH_only_vdw}) and (\ref{eq:KcRminrelation1}). The dotted vertical lines (black) are the same as those in \figref{VdWOnlyERmin}(a).}
\label{fig:aHHRminBoson}
\end{figure*}

To further elucidate the nature of this universality, we convert the horizontal axis into an authentic $s$-wave scattering length between the heavy atoms $a^{(\mathrm{HH})}_{\mathrm{with-dd}}$ in \figref{ModerateDipoleEaHHvHH_Boson}(g)-(l); in the presence of the dipole interaction, the $s$-wave scattering length between the heavy atoms $a^{(\mathrm{HH})}_{\mathrm{with-dd}}$ gets significantly modified from a purely van der Waals value of  $a^{(\mathrm{HH})}_{\mathrm{no-dd}}$ in \equref{aHH_only_vdw}~\cite{PhysRevA.64.022717,bohn2009quasi}. As the experimentally accessible scattering length between the Er atoms is $a^{(\mathrm{HH})}_{\mathrm{with-dd}}$ rather than $a^{(\mathrm{HH})}_{\mathrm{no-dd}}$, it should be more physical to use $a^{(\mathrm{HH})}_{\mathrm{with-dd}}$ to eliminate $R_{\mathrm{min}}$. In \figref{aHHRminBoson}, we show  $a^{(\mathrm{HH})}_{\mathrm{with-dd}}$ (circles) obtained by numerically solving the coupled-channel Schr\"{o}dinger equation between the two heavy atoms
\begin{equation}
 \label{eq:HeavyHeavy2bodyeq}  \left[-\frac{\nabla_{\bm{r}}^2}{M} -\frac{C_6}{r^6}+\frac{C_{dd}(1-3\cos^2\theta)}{r^3}\right]\psi(\bm{r})=E\psi(\bm{r}),
\end{equation}
with the boundary condition $\psi(|\bm{r}|=R_{\mathrm{min}})=0$, and compare it with $a^{(\mathrm{HH})}_{\mathrm{no-dd}}$. While they are almost indistinguishable for a weak dipole interaction (\figref{aHHRminBoson}(a)) except for the proximity of higher-partial wave induced resonances (dotted vertical lines)~\cite{PhysRevA.64.022717,PhysRevA.85.022703}, 
$a^{(\mathrm{HH})}_{\mathrm{with-dd}}$ is markedly different from $a^{(\mathrm{HH})}_{\mathrm{no-dd}}$ for a realistic dipole strength (\figref{aHHRminBoson}(b)), and stronger one (\figref{aHHRminBoson}(c)). This is expected because the realistic dipole strength of the Er atoms $a_{dd}/r_{\mathrm{vdw}}=0.86755....$ in \figref{aHHRminBoson}(b) (and stronger one in \figref{aHHRminBoson}(c)) is away from the perturbative regime which was confirmed to be $a_{dd}/r_{\mathrm{vdw}} \lesssim 0.5$ in \secref{res_weak}.

In \figref{ModerateDipoleEaHHvHH_Boson}(g)-(l), we show the binding energies of the Efimov states plotted against $a^{(\mathrm{HH})}_{\mathrm{with-dd}}$. Similar to \figref{ModerateDipoleEaHHvHH_Boson}(a)-(f), the data with different values of $R_{\mathrm{min}}$ mostly collapse into a single curve, demonstrating the universality. One notable difference is that the agreement between the numerical data and the purely van der Waals analytical curves (black) is much better in (g)-(l) than those in (a)-(f): except for the broad avoided crossing at $a_{\mathrm{with-dd}}^{(\mathrm{HH})}/r_{\mathrm{vdw}}\simeq 1$, a significant fraction of the numerical results agree well with the van der Waals curves not only for the weak dipole interaction, but also for the realistic dipole strength (\figref{ModerateDipoleEaHHvHH_Boson}(i)(j)) and stronger dipole interaction ((k)(l)). This suggests that the van der Waals QDT of the bosonic Efimov states in \equref{vdw_efimov_aHH_and_binding_energy} can capture the universal behavior of the Efimov states in the presence of the dipole interaction, once the effects of the dipole interaction are incorporated into the $s$-wave scattering length between the heavy atoms: $a^{(\mathrm{HH})}=a^{(\mathrm{HH})}_{\mathrm{with-dd}}$. In other words, the major role of the dipole interaction is to change, hence {\it renormalize}, the value of the $s$-wave scattering length between the heavy atoms, so that the van der Waals universality of the Efimov states in \equref{vdw_efimov_aHH_and_binding_energy} persists with the renormalized $s$-wave scattering length $a^{(\mathrm{HH})}_{\mathrm{with-dd}}$. This renormalized van der Waals universality for the dipolar bosonic Efimov states we have found here seems non-pertubative; the dipole interaction is so large that it significantly changes both the binding energies (\figref{ModerateDipoleEaHHvHH_Boson}(c)-(f)) and the  $s$-wave scattering length (\figref{aHHRminBoson}(b)(c))) from the purely van der Waals system. Some horizontal data points in \figref{ModerateDipoleEaHHvHH_Boson}(g)(l) which are not present in (a)-(f) are artifacts of the higher-partial-wave-induced resonances: the two-body problem between the heavy atoms in \equref{HeavyHeavy2bodyeq} not only shows the $s$-wave dominated resonances, but also shows the resonant enhancement of $a^{(\mathrm{HH})}$ induced by higher-partial-wave channels~\cite{PhysRevA.64.022717,PhysRevA.85.022703}, whose approximate positions are indicated by the vertical black dotted lines in \figref{aHHRminBoson}. The narrow nature of these higher-partial-wave-induced resonances leads to an abrupt change of $a^{(\mathrm{HH})}_{\mathrm{with-dd}}$ as  $R_{\mathrm{min}}$ is changed, resulting in flat energy spectra in \figref{ModerateDipoleEaHHvHH_Boson}(g)-(l). Equation~(\ref{eq:vdw_efimov_aHH_and_binding_energy}) cannot explain these flat data points. We note, however, that these flat energy spectra are absent in \figref{ModerateDipoleEaHHvHH_Boson}(a)-(b), and they should therefore be interpreted as artifacts originating from the two-body effect of $a^{(\mathrm{HH})}$ rather than a genuine three-body phenomenon. Physically, when we attempt to evaluate the binding energies of the Efimov states from an experimentally obtained value of $a^{(\mathrm{HH})}$ (see \secref{TBP_ErLi}), we can simply use the van der Waals curves of  \equref{vdw_efimov_aHH_and_binding_energy} by neglecting the stray flat spectra, except for the case where the two Er atoms are coincidentally in the proximity of the $s$-wave resonance induced by a $d$-wave or higher-angular-momentum molecular state. In the avoided crossing region, the variance of the spectrum is larger than in the other regions (see \figref{ModerateDipoleEaHHvHH_Boson}(l)). This is due to the finite values of $R_{\mathrm{min}}$, as it can be checked that reducing it further does lead to a converged universal behavior. For the finite values of $R_{\mathrm{min}}$ used in the figures (which is varied between 0.25 and 0.4 $r_{\mathrm{vdw}}$), it is still justified to call this region's behavior universal despite the noticeable energy variance because it remains much smaller than the Efimov period, in sheer contrast with the strength of the van der Waals interaction at short distance, which varies by a factor 16. Yet, a $d$-wave dimer of the two heavy atoms appearing around $a_{\mathrm{with-dd}}^{(\mathrm{HH})}/r_{\mathrm{vdw}}\simeq 1$ (see \figref{AppDimModBoson} in Appendix.~\ref{app:dimer} where the dimer energies are presented) seems to significantly affect two-body and three-body behaviors in this region, making the simple renormalized van der Waals universality of \equref{vdw_efimov_aHH_and_binding_energy} inadequate.

While the physical origin of the renormalized van der Waals universality is yet to be clarified, we suspect that the angular-momentum-insensitive nature of the short-range phase~\cite{PhysRevA.64.010701} is playing a crucial role; we have confirmed in our three-body calculations that a substantial fraction of the wave function is occupying $\ell \ge 2$ states. If the short-range phase shifts in various angular momentum channels are directly related with each other, as in the angular-momentun-insensitive QDT~\cite{PhysRevA.64.010701},  $a^{(\mathrm{HH})}$ characterizing the $s$-wave short-range phase can universally capture the scatterings occuring in all the angular momentum channels. We also suspect that a universal three-body repulsion appearing at moderately large distance may be another possible mechanism~\cite{PhysRevLett.108.263001,pascaleno3BP1,pascaleno3BP2,PhysRevLett.106.233201,PhysRevLett.107.233201}; in the three-body system of identical bosons interacting via the van der Waals interaction, the appearance of a non-adiabatic three-body repulsion at a large distance $r\simeq 2r_{\mathrm{vdw}} $ is ascribed to be the physical origin of the universality of the three-body parameter of the Efimov states~\cite{PhysRevLett.108.263001,pascaleno3BP1,pascaleno3BP2}. Furthermore, for three identical bosons interacting via purely dipolar interactions, a three-body repulsion has also been found to appear universally at $r\simeq a_{dd} $~\cite{PhysRevLett.106.233201}. The multi-channel effects with an interplay of the van der Waals and dipole interaction in our system may lead to a similar three-body repulsion scenario. We also remark that our universality of the Efimov states involving higher partial-wave channels is analogous to the partial-wave phase-locking mechanism found in the resonant exchange collisions~\cite{PhysRevLett.121.173401,PhysRevLett.121.173402}, where the phase shifts of different scattering channels between the atoms are related with each other.

\begin{figure}[!t]
	\centering
	\includegraphics[width=0.9\linewidth]{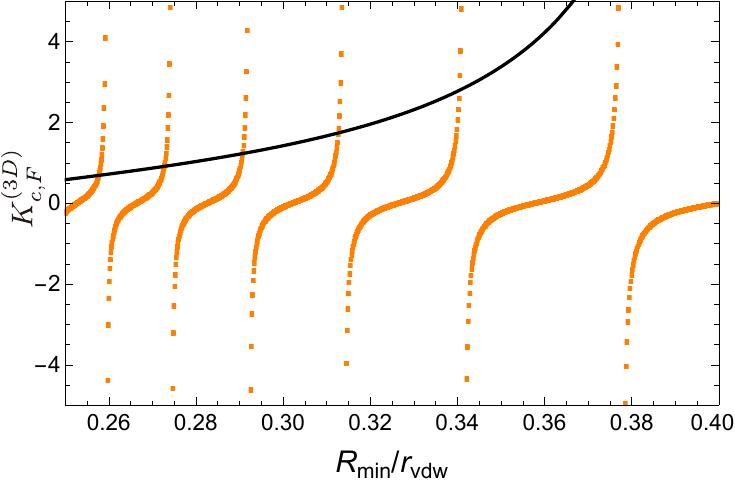} 
\caption{$K_{c,F}^{(3D)}$ between two heavy fermionic atoms for  $a_{dd}/r_{\mathrm{vdw}}=0.86755....$ corresponding to the realistic dipole strength between $^{167}$Er atoms ($a_{dd}=75.7a_0$ and $r_{\mathrm{vdw}} =65.9a_0$~\cite{PhysRevX.5.041029,Chomaz_2023}). The solid curve (black) is $K_{c,F}^{(3D)}$ calculated with purely dipole interaction (i.e., without the van der Waals interaction) in \equref{3DFermi_SchroEq}: $K_{c,F}^{(3D)} =J_3\left(q_{\mathrm{min}} \right)/ Y_3\left(q_{\mathrm{min}}\right)$ with $\displaystyle q_{\mathrm{min}}=\sqrt{\frac{48a_{dd}}{5R_{\mathrm{min}} }}$.}
\label{fig:Kc3D_RminFermi_v1}
\end{figure}

Since the $s$-wave scattering length between the heavy atoms characterizes the universal behavior of the bosonic Efimov states so precisely, it is tempting to use the $p$-wave scattering volume $v_p^{(\mathrm{HH})}$ to analyze the fermionic Efimov states. However,  $v_p^{(\mathrm{HH})}$ cannot be defined in the presence of the dipole interaction, because the dipole interaction is present even at large distance in the $\ell=1$ channel and the asymptotic wave function is no longer a free plane wave, in stark contrast to the $\ell=0$ channel~\cite{bohn2009quasi}. We therefore consider below an alternative scattering parameter $K_{c,F}^{(3D)}$ which characterizes the large-distance asymptotic wave function of the two heavy fermionic dipoles: we consider the Schr\"{o}dinger equation between the two heavy atoms
\begin{equation}
  \label{eq:3DFermi_SchroEq}
   \left[-\frac{1}{M} \frac{\partial^2}{\partial r^2}+\frac{2}{Mr^2}-\frac{C_6}{r^6}-\frac{4}{5}\frac{C_{dd}}{r^3}\right]u_{1}(r)=Eu_{1}(r).
\end{equation}
We have only kept the van der Waals interaction and the diagonal $\ell=1$ element of the dipole interaction, which are reasonably assumed to play major roles in determining the short- and long-range behaviors. At large distance, the van der Waals interaction is negligible in \equref{3DFermi_SchroEq}, so that the wave function behaves at vanishing energy as~\cite{PhysRevA.78.012702}
\begin{equation}
   \label{eq:3DFermiAsymptotocWF}
   u_{1}(r\rightarrow \infty)= \sqrt{r}\left[J_3\left({\sqrt{\frac{48}{5}\frac{a_{dd}}{r}}}\right)-K_{c,F}^{(3D)}Y_3\left({\sqrt{\frac{48}{5}\frac{a_{dd}}{r}}}\right)\right],
\end{equation}
where $J_3$ and $Y_3$ are the Bessel and Neumann functions corresponding to two independent solutions under the central $1/r^3$ potential. $K_{c,F}^{(3D)}$ is introduced in an analogous manner as the QDT parameter $K^c$ in the van der Waals case (see \equref{QDT_uKc_definition}), and hence is expected to be a universal parameter characterizing the short-range two-body phase shift. In \figref{Kc3D_RminFermi_v1}, we show $K_{c,F}^{(3D)}$ numerically calculated from Eqs.~(\ref{eq:3DFermi_SchroEq}) and (\ref{eq:3DFermiAsymptotocWF}) with the hard-wall boundary condition $u_{1}(r=R_{\mathrm{min}})=0$. We note that the van der Waals term is essential in determining the short-range phase: $K_{c,F}^{(3D)}$ is markedly different from that obtained by neglecting the van der Waals term (black solid), suggesting both the van der Waals and dipole interactions are relevant for characterizing the scattering between the two heavy fermionic atoms.

\begin{figure}[!t]
	\centering
	\includegraphics[width=1.0\linewidth]{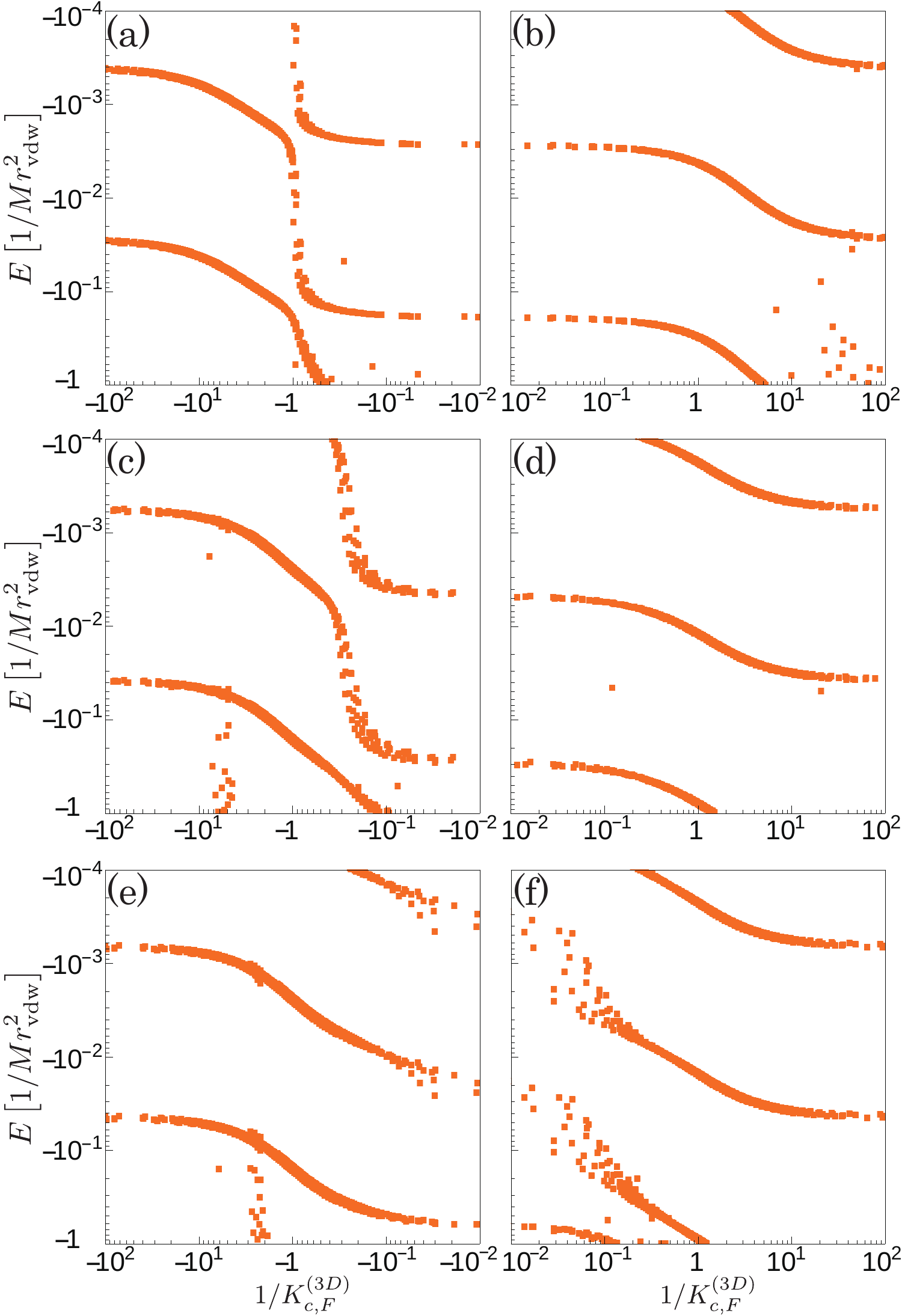} 
\caption{Binding energy of the fermionic Efimov states for the mass ratio  $M/m=27.7520...$ corresponding to $^{167}$Er-$^{167}$Er-$^6$Li system. $K_{c,F}^{(3D)}$ in the horizontal axis is obtained numerically from Eqs.~(\ref{eq:3DFermi_SchroEq}) and (\ref{eq:3DFermiAsymptotocWF}). Realistic dipole strength for $^{167}$Er-$^{167}$Er of $a_{dd}/r_{\mathrm{vdw}}=0.8705....$ (top row (a)(b)), and stronger dipole strengths of $a_{dd}/r_{\mathrm{vdw}}=1.6$ (middle row~(c)(d)), and $a_{dd}/r_{\mathrm{vdw}}=2.0$ (bottom row (e)(f)) are presented. }
\label{fig:ModerateDipoleEvHHK_Fermi}
\end{figure}

In \figref{ModerateDipoleEvHHK_Fermi}, we show the binding energies of the fermionic Efimov states plotted against $K_{c,F}^{(3D)}$ as obtained above. The numerical data of different short-range $R_{\mathrm{min}}$ values collapse into a single curve, demonstrating the universality of the fermionic Efimov states. The universality holds particularly well for a realistic $a_{dd}$ value of $^{167}$Er-$^{167}$Er (\figref{ModerateDipoleEvHHK_Fermi}(a)(b)), where only a few points of higher angular momentum character deviate from the universal curve. In contrast to the bosonic case, we cannot find an analytical expression which could explain the universal energy spectra owing to the impossibility of using  $v_p^{(\mathrm{HH})}$ and hence \equref{vdw_efimov_vHH_and_binding_energy}. The avoided crossing at $K_{c,F}^{(3D)}\simeq -1$ originates from the $\ell=3$ channel which is present without the dipole interaction at around $r_{\mathrm{vdw}}^3/v_p^{(\mathrm{HH})}\simeq -0.3$ (see \figref{VdWOnlyEaHHvHH}(c) and \figref{AppDimModFermion} in Appendix~\ref{app:dimer}). It shifts toward a larger negative $K_{c,F}^{(3D)}$ side with a decreased slope as the dipole strength is increased from a realistic value of $^{167}$Er-$^{167}$Er (\figref{ModerateDipoleEvHHK_Fermi}(a)(b)) toward larger values (\figref{ModerateDipoleEvHHK_Fermi}(c)-(f)). The foothill of the avoided crossing shows a larger variance for stronger dipole interactions (\figref{ModerateDipoleEvHHK_Fermi}(c)-(f)). This is likely be a precursor of the breakdown of our assumption that the single-channel short-range phase in $\ell=1$ universally characterizes the three-body system (see \secref{res_strong}), especially for a very strong dipole interaction and large $K_{c,F}^{(3D)}$ (\figref{ModerateDipoleEvHHK_Fermi}(c)(e)(f)).

Although the $K_{c,F}^{(3D)}$ parameter has the advantage of capturing the universal behavior of the fermionic Efimov states, it is marginally useful in predicting the three-body parameter of experimental systems of interest, e.g., $^{167}$Er-$^{167}$Er-$^6$Li. This is primarily because it is not easy, though not impossible, to determine the value of $K_{c,F}^{(3D)}$; it sensitively depends on the position (i.e. $R_{\mathrm{min}}$) and shape of the short-range repulsive potential, and it is therefore challenging to predict its value by theoretical calculations~\cite{PhysRevX.5.041029,PhysRevLett.109.103002,Kotochigova_2014}. Furthermore, as the $K_{c,F}^{(3D)}$ dependence appears as the next-next-next leading order if we perform the asymptotic expansion of the $J_3$ and $Y_3$ at large distance in \equref{3DFermiAsymptotocWF}, it contributes to the low-energy scattering cross section as a small correction term modifying the dominant quasi-universal contribution~\cite{bohn2009quasi}.




\subsection{\label{sec:res_strong}Strong dipole interaction $a_{dd}\gg r_{\mathrm{vdw}}$}

While the universal descriptions via partial-wave scattering parameters of $a^{(\mathrm{HH})}$ (boson) and  $K_{c,F}^{(3D)}$ (fermion) are shown in \secref{res_moderate} to be valid for moderate dipole strengths including the realistic values for Er atoms, they should break down in the limit of strong dipole interactions. This is because the single partial-wave description in defining the short-range two-body phase between the two heavy atoms becomes invalid since all the even (odd) angular momentum channels almost equally contribute for the bosons (fermions).

We therefore need to introduce an alternative scattering parameter which should characterize the short-range scattering phase of two strong dipoles. As the two dipoles prefer to be aligned in a head-to-tail, hence quasi-one-dimensional, configuration in the strong dipole limit, we consider a one-dimensional equation
\begin{equation}
  \label{eq:1D_SchroEq}
   \left[-\frac{1}{M} \frac{\partial^2}{\partial z^2}-\frac{C_6}{z^6}-\frac{2C_{dd}}{z^3}\right]u(z)=Eu(z),
\end{equation}
which corresponds to \equref{HeavyHeavy2bodyeq} constrained to the one-dimensional configuration on the $z$-axis. We introduce the scattering parameter $K_{c}^{(1D)}$ between the two dipoles inspired from the asymptotic behavior of the zero-energy wave function of \equref{1D_SchroEq} at $|z|\rightarrow \infty$~\cite{PhysRevLett.99.140406}
\begin{align}
   \label{eq:1DAsymptotocWF}
   u(z)\rightarrow (\mathrm{sgn}z)^P\sqrt{|z|} \Biggl[J_1\Biggl({\sqrt{\frac{24a_{dd}}{|z|}}}\Biggr) -K_{c}^{(1D)}Y_1\Biggl({\sqrt{\frac{24a_{dd}}{|z|}}}\Biggr)\Biggr],
\end{align}
where $J_1$ and $Y_1$ terms are two independent solutions of \equref{1D_SchroEq} without the van der Waals interaction. We impose a hard-wall boundary condition $u(z = \pm R_{\mathrm{min}})=0$ corresponding to the three-dimensional counterpart. Since the positive and negative $z$ regions are separated, the bosonic and fermionic systems become equivalent, with the same value of $K_{c}^{(1D)}$; their only difference appears in the overall parity, i.e. $P=0$ for bosons and $P=1$ for fermions in \equref{1DAsymptotocWF}.

\begin{figure}[!t]
	\centering
	\includegraphics[width=1.02\linewidth]{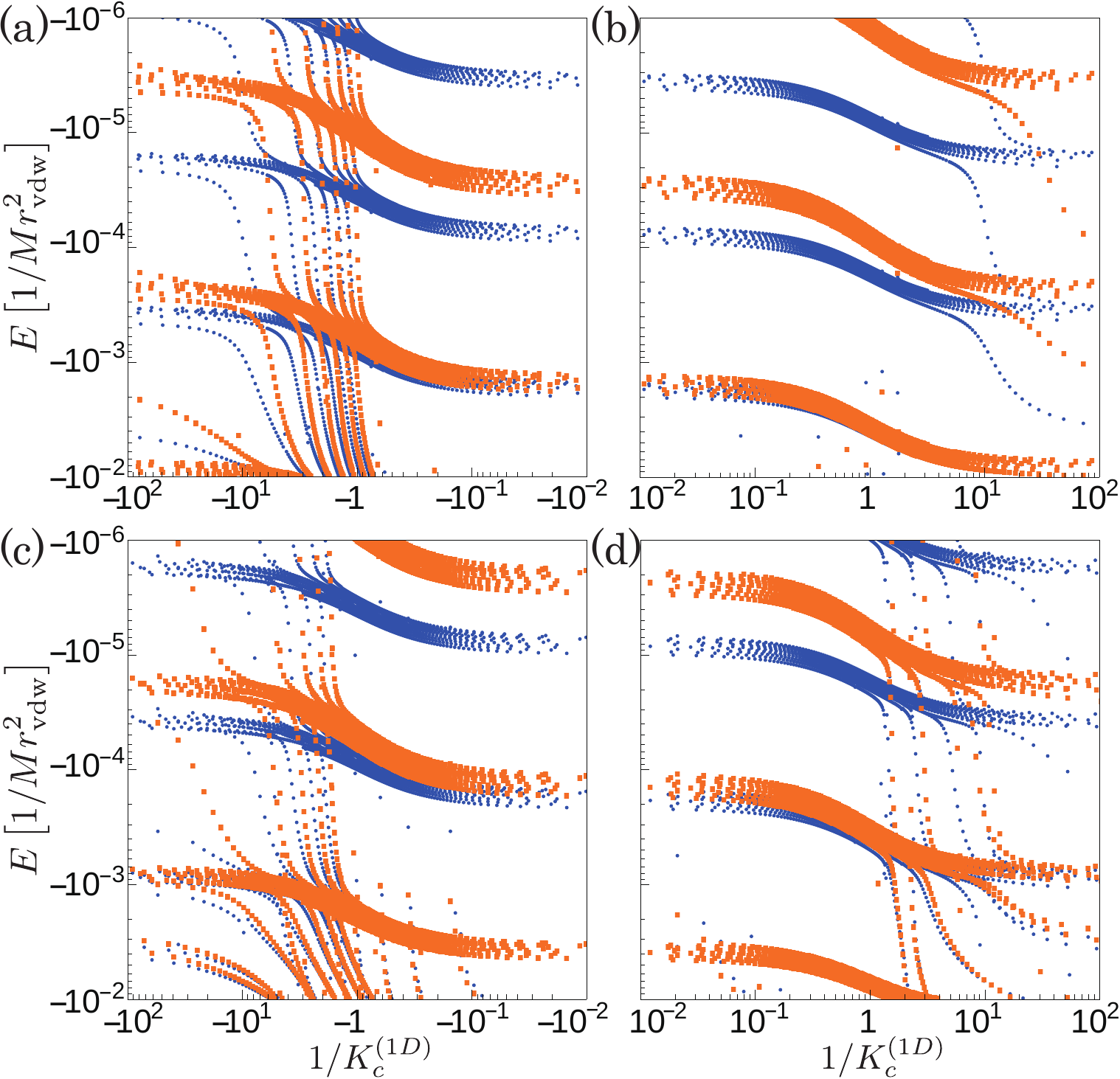} 
\caption{Binding energy of the bosonic (blue circles) and fermionic (orange squares) Efimov states for strong dipole interactions of $a_{dd}/r_{\mathrm{vdw}}=50$ (top row (a)(b)), and $a_{dd}/r_{\mathrm{vdw}}=100$ (bottom row (c)(d)). $K_{c}^{(1D)}$ in the horizontal axis is obtained numerically from Eqs.~(\ref{eq:1D_SchroEq}) and (\ref{eq:1DAsymptotocWF}). The mass ratio is taken as $M/m=27.5855...$, which corresponds to $^{166}$Er-$^6$Li for the boson, while for the fermion it is hypothetically taken to be the same value as the boson to demonstrate the difference via the statistics of the particles.}
\label{fig:StrongDipoleEK}
\end{figure}

In \figref{StrongDipoleEK}, the binding energies of the Efimov states obtained from \equref{HeavyHeavy2bodyeq} are plotted as a function of $K_{c}^{(1D)}$ in Eqs.~(\ref{eq:1D_SchroEq}) and (\ref{eq:1DAsymptotocWF}). The bosonic (blue circles) and fermionic (orange squares) systems, hypothetically assumed to be the same mass ratio, show rather similar behaviors, in stark contrast to the weak and moderate dipole systems of Secs.~\ref{sec:res_vdw}-\ref{sec:res_moderate}. This is because the parity of the wave function plays a marginal role when the two heavy atoms are mostly aligned in a quasi-one-dimensional configuration along the $z$-axis, and the $z>0$ and $z<0$ regions are separated by the hard-wall boundary condition. This is particularly true when the size of the trimer is much smaller than $a_{dd}$, beyond which the centrifugal repulsion in \equref{HeavyHeavy2bodyeq} representing the three-dimensional nature becomes more dominant than the dipole term. For the tightly bound states in \figref{StrongDipoleEK}, the bosonic and fermionic systems become almost indistinguishable, supporting this scenario. In other words, the statistics of the particles is irrelevant for the strongly dipolar Efimov states as long as their binding energy is larger than the dipole scale $1/Ma_{dd}^2$.

\begin{figure}[!t]
	\centering
	\includegraphics[width=1.02\linewidth]{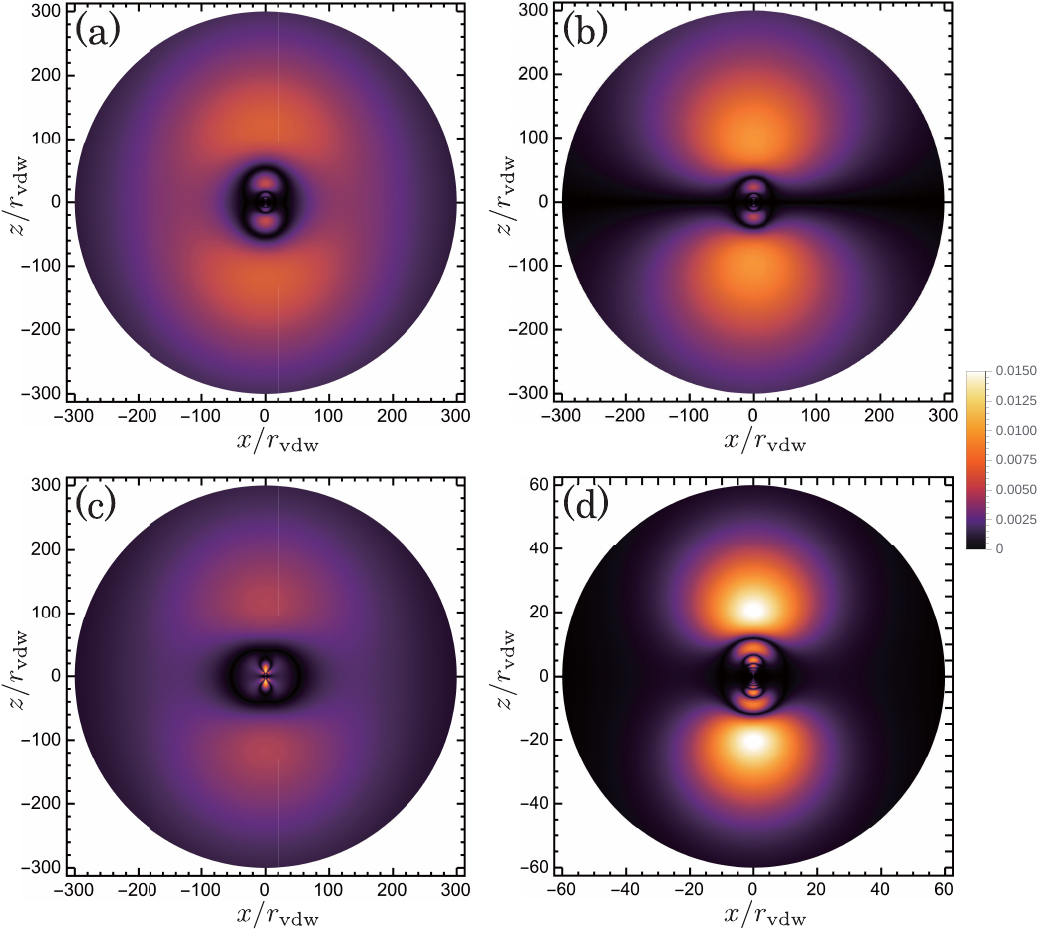} 
\caption{Contour plot of $\displaystyle |r \psi(\bm{r})|$ of the relative position of the two heavy particles in the Efimov states, plotted in the $y=0$ plane. (a) Shallow bound states in the universal regime for the bosons, and (b) for the fermions. (c) Shallow bound states in the non-universal regime (i.e, a steep slope region between the avoided crossings in \figref{StrongDipoleEK}(a)) for the bosons, and (d) tightly bound states for the bosons (plotted in a smaller $xz$ region than (a)-(c)). For all (a)-(d), the dipole strength and mass ratio are the same as those in \figref{StrongDipoleEK}~(a)(b), with the following parameters: (a) $Mr_{\mathrm{vdw}}^2|E| =1.24\times 10^{-4} $, $(K_{c}^{(1D)})^{-1}=0.71...$ ($R_{\mathrm{min}}/r_{\mathrm{vdw}}=0.25225...$),  (b) $Mr_{\mathrm{vdw}}^2|E| = 1.23\times 10^{-4} $, $(K_{c}^{(1D)})^{-1}=2.58...$ ($R_{\mathrm{min}}/r_{\mathrm{vdw}}=0.2617...$), (c) $Mr_{\mathrm{vdw}}^2|E| =1.10\times 10^{-4}  $, $(K_{c}^{(1D)})^{-1}=-1.14...$ ($R_{\mathrm{min}}/r_{\mathrm{vdw}}=0.25758...$), and  (d) $Mr_{\mathrm{vdw}}^2|E| =9.72\times 10^{-3}  $, $(K_{c}^{(1D)})^{-1}=-3.42...$ ($R_{\mathrm{min}}/r_{\mathrm{vdw}}=0.2592...$).
 }
\label{fig:StrongDipoleWFplot}
\end{figure}

The data points with a small slope in \figref{StrongDipoleEK}, including the weakly bound states, tend to collapse into a single curve for both the bosonic and fermionic systems. Although the spreading of the data points is larger than those in Figs.~\ref{fig:ModerateDipoleEaHHvHH_Boson} and~\ref{fig:ModerateDipoleEvHHK_Fermi}, it remains much smaller than the Efimov period, even with an almost 16 times change in the short-range part of the van der Waals interaction. We have also confirmed that the spreading gets smaller as $R_{\mathrm{min}}$ is decreased, in a similar manner to Fig.~\ref{fig:ModerateDipoleEaHHvHH_Boson}. We can thus conclude that $K_{c}^{(1D)}$ universally characterizes the short-range phase, hence the Efimov states with the strong dipole interactions. On the other hand, the data points with steeper slopes do not seem to show the universality with respect to $K_{c}^{(1D)}$. Since those states have significant contributions from large angular momentum states $\ell \gg 1$ and the angular insensitive nature of their short-range phase can break down, it is not unexpected to find that their features cannot be universally captured by the one-dimensional scattering model in Eqs.~(\ref{eq:1D_SchroEq}) and (\ref{eq:1DAsymptotocWF}) where the angular momentum is absent (see also dimers in \figref{AppDimStrong} in Appendix~\ref{app:dimer} with similar behaviors). Nevertheless, the physical mechanism behind the universal and non-universal behaviors in \secref{res_moderate} and \figref{StrongDipoleEK} and how they change between them have yet to be clarified.

In \figref{StrongDipoleWFplot}, we show the amplitude of the wave function. For a bosonic Efimov state with a small binding energy (\figref{StrongDipoleWFplot}~(a)), it essentially comprises two regions; the inner region $R\lesssim a_{dd}$ where the dipole interaction is so strong that the two heavy particles are preferentially aligned on the $z$ axis, and the outer region $R\gtrsim a_{dd}$ where the dipole interaction is smaller than the kinetic term, thus forming a three-dimensional halo state of a predominantly $s$-wave nature. They are neither a perfect one-dimensional geometry nor a spherical $s$-wave halo owing to the strong admixture of angular momentum states, but their universal nature is essentially determined by the phase in the short-range region, where the one-dimensional configuration is dominant and hence $K_{c}^{(1D)}$ is a universal parameter.  For fermionic Efimov states with a similarly small value of the binding energy (\figref{StrongDipoleWFplot}~(b)), the short-range and long-range natures are almost the same, with the only notable difference being the presence of a node in the wave function at $z=0$, representing the anti-symmetrization.  In contrast, the wave function in the non-universal regime (\figref{StrongDipoleWFplot}~(c)) shows a distinct feature: the wave function in the inner region $R\lesssim a_{dd}$ no longer shows a simple quasi-one-dimensional character, but rather a strongly anisotropic shape, which is likely induced by an admixture of various angular momentum states. This is consistent with the breakdown of the universality with respect to $K_{c}^{(1D)}$. When the binding energy is so large that the size of the wave function is smaller than $a_{dd}$ (\figref{StrongDipoleWFplot}~(d)), the whole wave function is localized in the inner region $R\lesssim a_{dd}$, forming an almost perfect quasi-one-dimensional state. In this limit, the hard-wall boundary condition $R= R_{\mathrm{min}}$ necessarily produces a node in the wave function at $z\approx 0$, so that the bososnic and fermionic wave functions become  the same except for the sign change. In this limit, the bosonic and fermionic systems show the same universal behaviors. While this limit of boson-fermion duality, analogous to the Tonks-Girardeau limit~\cite{PhysRev.50.955,10.1063/1.1703687}, is fascinating, it is challenging to achieve because it requires a very large dipole interaction strength, which is impossible for cold atoms with magnetic dipole interactions, but may be possible with polar molecules~\cite{ni2008high,cornish2024quantum} or Rydberg atoms~\cite{RevModPhys.82.2313,GALLAGHER2008161}.

\section{\label{sec:TBP_ErLi}Three-body parameters of Er-Er-Li Efimov states}
 In \tabref{result_list_TBP}, we list some combinations of isotopes of the Er-Li systems, together with their physical parameters. The dipole lengths are comparable to the Er-Er van der Waals length $a_{dd}/r_{\mathrm{vdw}} \simeq 0.87$, which lie outside the perturbative regime $a_{dd}/r_{\mathrm{vdw}} \lesssim 0.5$ discussed in \secref{res_weak}. Nevertheless, they are not strong enough to be in the strong-dipole limit discussed in \secref{res_strong}; they are in the parameter region well described by the renormalized van der Waals universality shown in \secref{res_moderate}. We can therefore use Eq.~(\ref{eq:vdw_efimov_aHH_and_binding_energy}) to evaluate the binding energy of the Efimov states for the bosonic Er atoms interacting with a $^6$Li atom in unitary limit. The $s$-wave scattering length between the Er atoms $a^{(\mathrm{HH})}$ generally depends on the applied magnetic field, owing to the existence of a series of broad and narrow Feshbach resonances~\cite{PhysRevX.5.041029,PhysRevLett.109.103002}, and it is yet to be determined by experiments or by sophisticated theoretical calculations. Still, by using the background $s$-wave scattering lengths reported in the literature~\cite{PhysRevX.5.041029,Chomaz_2023} as representative values, we can estimate the binding energy of the Efimov states $|E| = \kappa_*^2 / M $. The results are tabulated in \tabref{result_list_TBP}; we find that the difference in the values of $a^{(\mathrm{HH})}$ significantly affects the three-body parameter $\kappa_*$ across the isotopes while the mass and $a_{dd}$ are almost the same and therefore play marginal roles. The values of $\kappa_*$ in \tabref{result_list_TBP} can be readily improved with better knowledge of the Er-Er scattering length $a^{(\mathrm{HH})}$. We also note that the larger binding energy states in \tabref{result_list_TBP} may not be so accurate because we use the low-energy analytical formula in Eq.~(\ref{eq:vdw_efimov_aHH_and_binding_energy}); this can also be straightforwardly improved using the numerical results of \secref{res_moderate}.

\begin{table}[!t]
\caption{Three-body parameters of the Er-Er-Li Efimov states: binding energy at the unitary Er-Li scattering length $|E|= \kappa_*^2 / M$, and the Er-Li scattering length $a_-^{(\mathrm{HL})}$ ($a_*^{(\mathrm{HL})}$) at which the Efimov states dissociate into three atoms (an atom and a dimer), respectively. The van der Waals and dipole lengths between the Er atoms are from Refs~\cite{PhysRevX.5.041029,Chomaz_2023} with the mass scaling. $a^{(\mathrm{HH})}$ between the heavy atoms are estimated by the background scattering lengths $a_{bg}$ tabulated in Refs.~\cite{Chomaz_2023}. $a_0$ is the Bohr radius.}
\centering
\label{tab:result_list_TBP} 
\begin{tabular}{ccccccc }
\hline
Species & $r_{\mathrm{vdw}}$[$a_0$]& $a_{dd}$[$a_0$] & $a^{(\mathrm{HH})}$ [$a_0$] & $\kappa_* r_{\mathrm{vdw}} $ & $\dfrac{a_-^{(\mathrm{HL})}}{r_{\mathrm{vdw}}}$& $\dfrac{a_*^{(\mathrm{HL})}}{r_{\mathrm{vdw}}}$\\
 \hline
 \multirow{4}{*}{$^{166}$Er-$^{6}$Li}& \multirow{4}{*}{75.5} & \multirow{4}{*}{65.5} & \multirow{4}{*}{68} &  0.495 & -10.1 & 0.269\\
 & & & & 0.107 & -46.7 & 1.25\\
 & & & & 2.29$\times 10^{-2}$ & -217 & 5.81\\
 & & & & 4.94$\times 10^{-3}$ & -1.01$\times 10^{3}$ & 27.0\\
 \hline
 \multirow{4}{*}{$^{168}$Er-$^{6}$Li}& \multirow{4}{*}{75.8} & \multirow{4}{*}{66.3} & \multirow{4}{*}{137} & 0.352&-14.2 & 0.381\\
 & & & & 7.65$\times 10^{-2}$& -65.3 & 1.75\\
 & & & & 1.66$\times 10^{-2}$& -300& 8.05\\
 & & & & 3.62$\times 10^{-3}$&  -1.38$\times 10^{3}$ & 37.0\\
 \hline
 \multirow{4}{*}{$^{170}$Er-$^{6}$Li}& \multirow{4}{*}{76.0} & \multirow{4}{*}{67} & \multirow{4}{*}{221} & 0.298& -16.8 & 0.452\\
 & & & & 6.55$\times 10^{-2}$& -76.6& 2.06\\
 & & & & 1.44$\times 10^{-2}$& -349& 9.37\\
 & & & & 3.16$\times 10^{-3}$& -1.59$\times 10^{3}$& 42.7\\
  \hline
\end{tabular}
\end{table} 
 
  While there are spectroscopic methods to directly associate the Efimov trimers and observe their binding energies~\cite{LompeScience11,PhysRevLett.106.143201,PhysRevLett.108.210406,PhysRevLett.122.200402}, it is challenging to perform such experiments, especially in the unitary limit. It is more straightforward to observe the three-body loss rate of the atoms for variable $s$-wave scattering lengths $a^{(\mathrm{HL})}$, and search for the negative scattering length $a=a_-^{(\mathrm{HL})}$ at which the loss rate peak appears~\cite{kraemer2006evidence,PhysRevLett.112.250404,PhysRevLett.113.240402,PhysRevLett.115.043201,PhysRevLett.103.130404,PhysRevLett.107.120401}. Alternatively, one can prepare a mixture of heavy-light Feshbach molecules and heavy atoms, and observe the dimer-atom loss rate to find the positive scattering length $a=a_*^{(\mathrm{HL})}$ at which its peak appears~\cite{PhysRevLett.105.023201}. The values $a_-^{(\mathrm{HL})}$ and $a_*^{(\mathrm{HL})}$ correspond to the points at which the Efimov states dissociate into three atoms, or a dimer and an atom, respectively. Notably, any of $a_-^{(\mathrm{HL})}$, $a_*^{(\mathrm{HL})}$, or $\kappa_*$ can be regarded as the universal scale characterizing the Efimov states because they can be related with each other by the universal Efimov theory~\cite{efimov1970energy,efimov1973energy,naidon2017efimov,RevModPhys.89.035006,d2018few,Braaten2006259}. Using the zero-range Efimov theory, we have estimated $a_-^{(\mathrm{HL})}$, $a_*^{(\mathrm{HL})}$ from the $\kappa_*$ values. The results are presented in the last two columns of \tabref{result_list_TBP}. The values of $a_-^{(\mathrm{HL})}$ corresponding to the most tightly bound Efimov states are likely to be either around $a_-^{(\mathrm{HL})}/r_{\mathrm{vdw}}\approx$ $-10$-$17$, or adjacent ones with $\approx -2$-$4$ (not tabulated). With the low-energy formula Eq.~(\ref{eq:vdw_efimov_aHH_and_binding_energy}) and the zero-range theory, we cannot conclude which of them are the ground Efimov state. We also note that they can be significantly affected by finite-range effects owing to their relatively large binding energy, while results for the more weakly bound states in \tabref{result_list_TBP} should be more quantitatively reliable. Similarly, we expect that the dissociation points of the excited Efimov states around $a_*^{(\mathrm{HL})}/r_{\mathrm{vdw}}$$\approx5$-$9$ are more likely to be experimentally observed than those of $\approx 1-2$ or $\approx 0.2-0.4$ corresponding to the tightly bound states. This is analogous to the measurement of $a_*$ in systems of three identical bosons, for which the loss-rate peak corresponding to the excited Efimov states appear clearly, whereas those for the ground and first excited states may not appear, or only barely~\cite{PhysRevA.95.032707,PhysRevA.99.012702}.

  For the fermionic Efimov states explored in recent cold-atom experiments of $^{167}$Er-$^{6}$Li mixture~\cite{ErLiFR2}, the shortage of scattering data between the fermionic $^{167}$Er atoms prevents us from making a similar estimate. Since the fermionic Efimov states show the universal behavior characterized by $K_{c,F}^{(3D)}$ (see \secref{res_moderate}), we can make a theoretical prediction once the value of $K_{c,F}^{(3D)}$ is known. However, since the low-energy scattering between the fermionic $^{167}$Er atoms is dominated by the universal dipolar scattering, the determination of $K_{c,F}^{(3D)}$ should be challenging, requiring the determination of the sub-leading contributions in the $^{167}$Er-$^{167}$Er scattering.

\section{\label{sec:Concl}Conclusion \& Outlook}
We have studied the universality of the Efimov states in a mass-imbalanced three-body system of heavy-heavy-light atoms with a unitary interaction between the heavy-light atoms, with the major motivation of elucidating the interplay between the van der Waals and magnetic dipole interactions in Er-Er-Li Efimov states. By reducing the three-body problem to a coupled-channel two-body equation between the heavy atoms using the Born-Oppenheimer approximation, we have investigated analytically and numerically the three-body parameters (i.e., binding energies in the unitary limit) of the Efimov states for various strengths of dipole interactions between the bosonic and fermionic heavy atoms.

In the absence of dipole interaction, using the quantum defect theory, we derive analytical formulas of the three-body parameters of the Efimov states, which are represented as explicit universal functions of the van der Waals length and $s$-wave scattering length ($p$-wave scattering volume) between the heavy bosonic (fermionic) atoms. The analytical formulas excellently reproduce the numerical results reported in Ref.~\cite{PhysRevA.95.062708}, demonstrating the van der Waals universality in an analytical manner.

In the presence of dipole interaction, different orbital angular momentum states get coupled, thus we have performed coupled-channel numerical calculations. For a weak dipole interaction, we find that the three-body parameters are affected by the dipole interaction in a different manner for the bosons and fermions. Due to the presence (absence) of the dipole matrix element for bosons in the $L=0$ (fermions in $L=1$) states, the three-body parameter gets a quadratic (linear) shift with respect to a perturbative dipole interaction for the bosons (fermions), respectively.

When the dipole interaction is as strong as the van der Waals interaction, we find that the values of the three-body parameters get significantly altered from those of a non-dipolar system. However, once the $s$-wave scattering length between the heavy atoms is calculated in the presence of both the dipole and van der Waals interactions, the QDT formulas derived for a pure van der Waals system can well reproduce the three-body parameters of the bosonic Efimov states. Using this {\it renormalized} van der Waals universality, we have estimated the three-body parameters of the Er-Er-Li Efimov states for the bososnic Er isotopes.  For the fermionic heavy atoms, on the other hand, the $p$-wave scattering volume can no longer be defined, owing to the long-range nature of the dipole interaction, so that the renormalized van der Waals universality does not hold. However, they can still be described universally with a low-energy scattering parameter between the two fermionic dipoles which characterizes the phase shift of an asymptotic $p$-wave state, so that the universality of the three-body parameter holds true.

For an extremely large dipole interaction, neither the $s$-wave scattering length nor the $p$-wave low-energy scattering parameter can universally characterize the Efimov states, because the two dipoles get preferentially aligned in a linear geometry at short distance and they are neither $s$-wave nor $p$-wave dominant. In this limit, a quasi-one-dimensional scattering parameter is capable of capturing the short-range scattering phase. We show that except for some regions between avoided crossings, the Efimov states can be universally described by the one-dimensional scattering parameter.

With the above studies, we have elucidated how the interplay of the van der Waals and dipole interactions can affect the universality of the Efimov states. In particular, we have clarified the roles of the statistics of the particles and the angular momentum channels, in the presence of both isotropic and anisotropic forces. Our work is not only relevant for the current challenge of observing Efimov states in the highly mass-imbalanced cold atom mixture of Er-Li atoms~\cite{ErLiFR1,ErLiFR2}, but also can shed new light on halo nuclear phenomena. The Efimov states are believed to appear universally in various strongly interacting quantum systems~\cite{kunitski2015observation,nishida2013efimov,naidon2017efimov}. In particular, Efimov-like weakly bound three-body states have been actively studied in nuclear systems, such as neutron-rich nuclei~\cite{AnnRev_HamPlatt,PhysRevLett.120.052502,hammer2017effective,PhysRevC.100.011603}, the Hoyle state of $^{12}$C~\cite{hoyle1954nuclear,higa2008alphaalpha}, and in near-breakup excited states of nuclei~\cite{endotanaka2023}. In nuclear physics, the tensor force has been pointed out to play a crucial role in determining the stability and reactions of nuclei. Notably, the tensor force has an anisotropic nature similar to the dipole-dipole interaction~\cite{RingShuck}. Investigating the interplay of the van der Waals interaction and dipole interaction is thus analogous to the nuclear problems where the interplay of the central part of the strong force and the tensor force is relevant. Our study of the dipole and van der Waals interactions in strongly interacting quantum few-body systems lays the basis for the quantum simulation of nuclear phenomena using dipolar cold-atom mixtures~\cite{ErLiFR1,ErLiFR2,CrLiFR,Chomaz_2023,schafer2020tools}.

\begin{acknowledgments}
 We thank Yoshiro Takahashi, Yousuke Takasu, Lucas Happ, and Emiko Hiyama for fruitful discussions. This work was supported by JSPS KAKENHI Grants No. JP22K03492, JP23H01174, and JP23K03292. K.O. acknowledges support from Graduate Program on Physics for the Universe (GP-PU) of Tohoku University. S.E. acknowledges support from the Institute for Advanced Science, University of Electro-Communications.
\end{acknowledgments}

\appendix
\section{\label{app:vdw2body} Analytical Solution of \equref{BOeq_vdwonly}}
 Equation~(\ref{eq:BOeq_vdwonly}) can be analytically solved in a similar manner as Ref.~\cite{PhysRevA.58.1728}: introducing $x  \equiv2\left(\frac{r_{\rm vdw}}{r}\right)^2$, $\Delta  \equiv Mr_{\rm vdw}^2 E/4$ and $u_{\ell}(r)  \equiv\sqrt{r} \mathcal{U}_{\ell}(x)$,  \equref{BOeq_vdwonly} is written in the dimensionless form as
 \begin{equation}
   \label{eq:sch_vdw_eq_radial_non_dim}
   \left[x^2\frac{d^2}{dx^2}+x\frac{d}{dx}+x^2-\nu_0^2\right] \mathcal{U}_{\ell}(x)=-\frac{2\Delta}{x} \mathcal{U}_{\ell}(x),
\end{equation}
where $\nu_0 = s_\ell/ 2$. In the absence of the induced attraction via the light particle,  \equref{BOeq_vdwonly} is the van der Waals potential problem; the attraction modifies the last term in the left-hand side from the quarter integer $\nu_0 = (2\ell +1)/4$ into $\nu_0 =s_\ell/ 2 $. When the Efimov effect is present (absent) as in the $\ell=0$ (large $\ell$) channel, $\nu_0$ is purely imaginary (real).

In a similar manner as Ref.~\cite{PhysRevA.58.1728}, the two linearly independent solutions of \equref{BOeq_vdwonly} are represented as
 \begin{align}
 \label{eq:fgbar_BesselSeries}
 \begin{split}
   \overline{f_{\ell}}(r)&=\sqrt{r}\sum_{n}b_n J_{\nu+n}\left(x\right), \\
   \overline{g_{\ell}}(r)&=\sqrt{r}\sum_{n}b_n Y_{\nu+n}\left(x\right).
   \end{split}
\end{align}
$b_n$ and $\nu$ are determined from the same transcendental equations in Ref.~\cite{PhysRevA.58.1728}, with the modification $\nu_0 = (2\ell +1)/4 \rightarrow \nu_0 =s_\ell/ 2 $. In order to have favorable short-range asymptotic features, it is convenient to consider $f_{\ell}^c$ and $g_{\ell}^c$, which are defined as linear transforms of $\overline{f_{\ell}}$ and  $\overline{g_{\ell}}$ as
\begin{small}
 \begin{align}
 \begin{split}
  f_{\ell}^c(r)&=\frac{1}{\sqrt{2(X_\ell^2+Y_\ell^2)}}\left[\cos\left(\frac{\nu \pi}{2} + \theta_\ell \right)\overline{f_{\ell}}-\sin\left(\frac{\nu \pi}{2} + \theta_\ell \right)\overline{g_{\ell}}\ \right], \\
 g_{\ell}^c(r)&=\frac{-1}{\sqrt{2(X_\ell^2+Y_\ell^2)}}\left[\sin\left(\frac{\nu \pi}{2} + \theta_\ell \right)\overline{f_{\ell}}+\cos\left(\frac{\nu \pi}{2} + \theta_\ell \right)\overline{g_{\ell}}\ \right],
 \end{split}
\end{align}
\end{small}
where $\tan \theta_\ell = Y_{\ell}/X_\ell$ with $X_\ell = \sum_n (-1)^n b_{2n}$ and $Y_\ell = \sum_n (-1)^n b_{2n+1}$.

The short-distance asymptotic forms of $f_{\ell}^c$ and  $g_{\ell}^c$ for $r\ll r_{\rm vdw}$ are 
\begin{align}
   \label{eq:fc_gc_shortrange}
   \begin{split}
    & f^c_{\ell}(r)\rightarrow \frac{1}{\sqrt{2\pi}} \frac{r^{3/2}}{r_{\rm vdw}} \cos\left[x-\frac{\pi}{4}\right], \\
    & g^c_{\ell}(r)\rightarrow -\frac{1}{\sqrt{2\pi}} \frac{r^{3/2}}{r_{\rm vdw}} \sin\left[x-\frac{\pi}{4}\right].
\end{split}
\end{align}
Notably, they are independent of angular momentum nor energy, which are essential features for the angular-momentum-insensitive QDT~\cite{PhysRevA.64.010701}.

The large-distance asymptotic forms of $f_{\ell}^c$ and  $g_{\ell}^c$ for $E=k^2/M>0$ are
\begin{equation}
   \label{positive_energy_fc_gc_long_range_asympt}
   \left[
      \begin{array}{c}
         f^c_{\ell} \\
         g^c_{\ell}
      \end{array}
      \right]\to
   \sqrt{ \frac{2}{\pi k} }
   \left[
      \begin{array}{cc}
         Z^c_{fs} & Z^c_{fc} \\
         Z^c_{gs} & Z^c_{gc}
      \end{array}
      \right]
   \left[
      \begin{array}{c}
         \sin\left(kr-\frac{l\pi}{2}\right) \\
         -\cos\left(kr-\frac{l\pi}{2}\right)
      \end{array}
      \right],
\end{equation}
and for $E= - \kappa^2/M<0$ are
\begin{equation}
   \label{negative_energy_fc_gc_long_range_asympt}
   \left[
      \begin{array}{c}
         f^c_{\ell} \\
         g^c_{\ell}
      \end{array}
      \right]\to
   \sqrt{ \frac{1}{2\pi \kappa} }
   \left[
      \begin{array}{cc}
         W^c_{f+} & W^c_{f-} \\
         W^c_{g+} & W^c_{g-}
      \end{array}
      \right]
   \left[
      \begin{array}{c}
         -2e^{-\kappa r} \\
         e^{\kappa r}
      \end{array}
      \right].
\end{equation}
The elements of the matrices $Z^c$ and $W^c$ are the same as those in Ref.~\cite{PhysRevA.58.1728}, with the modification $\nu_0 = (2\ell +1)/4 \rightarrow \nu_0 =s_\ell/ 2 $.

At vanishing energy $\Delta \rightarrow 0$, the wave function is dominantly described by the single term of the Bessel series in \equref{fgbar_BesselSeries} and $\nu \simeq \nu_0$:
\begin{equation}
   \begin{split}
      \label{eq:vdw_only_0_energy_sol}
      f^c_{\ell}(r)&=\sqrt{\frac{r}{2}}\left[\cos\frac{\pi \nu_0}{2}J_{\nu_0}\left(x\right)-\sin\frac{\pi \nu_0}{2}Y_{\nu_0}\left(x\right)\right] \\
      g^c_{\ell}(r)&=\sqrt{\frac{r}{2}}\left[-\sin\frac{\pi \nu_0}{2}J_{\nu_0}\left(x\right)-\cos\frac{\pi \nu_0}{2}Y_{\nu_0}\left(x\right)\right].
   \end{split}
\end{equation}
In the absence of the light particle, $\nu_0 = (2\ell +1)/4$ and the wave function is essentially the Bessel function with a quarter integer index as found in Ref.~\cite{PhysRevA.48.546}. In the presence of the light particle and hence the Efimov effect, the wave function is still the Bessel function, but now with a pure-imaginary index $\nu_0 =s_\ell/ 2 $; this reflects the log-periodic behavior of the Efimov states.  Using $Y_{\alpha}(z) = (J_\alpha (z) \cos \alpha \pi - J_{-\alpha}(z))/\sin \alpha \pi$ and $J_{-i q}(r)= \left(J_{i q}(r)\right)^*$ for real $q$ and $r$,  \equref{vdw_only_0_energy_sol} is rewritten as
\begin{equation}
   \label{eq:fc_gc_EfimovReImWF_app}
   \left[
      \begin{array}{c}
         f^c_{\ell} \\
         g^c_{\ell}
      \end{array}
      \right] = 
   \sqrt{ \frac{r}{2} }
   \left[
      \begin{array}{c}
         \mathrm{Re}\left[J_{\frac{s_\ell}{2}}\left(x \right)\right] / \cosh \frac{\pi |s_\ell|}{4}\\
         -\mathrm{Im}\left[J_{\frac{s_\ell}{2}}\left(x \right)\right] / \sinh \frac{\pi |s_\ell|}{4}
      \end{array}
      \right].
\end{equation}

\section{\label{app:QDTexpansion} Low-energy QDT expansion}
Following Refs.~\cite{BoGao2004,PhysRevA.80.012702}, we can perform an expansion (QDT expansion) at low energy. Indeed,  when $|\Delta|\ll1$, 
\begin{equation}
\label{eq:nu_QDT_expansion}   \nu = \nu_0 - \frac{3}{2\nu_0}\frac{\Delta^2}{(\nu_0^2-1)(4\nu_0^2-1)} + O(\Delta^4),
\end{equation}
from which we find $X_{\ell} \simeq1$ and 
\begin{equation}
\label{eq:theta_QDT_expansion}    \tan\theta_\ell\simeq Y_\ell \simeq -\frac{4\Delta}{(4\nu_0^2-1)}.
\end{equation}
We also obtain 
\begin{equation}
   \begin{split}
      \label{eq:M_low_energy}
      M_{\ell}&\simeq|\Delta|^{2\nu_0}\frac{2\pi^2}{[\Gamma(\nu_0)\Gamma(1+2\nu_0)]^2}\frac{1}{\sin(\pi\nu_0)\sin(2\pi\nu_0)},
   \end{split}
\end{equation}
where $M_{\ell}\equiv G_{\ell}(-\nu)/ G_{\ell}(\nu)$ with $G_\ell$ is defined as
\begin{equation}
G_\ell(\nu) \equiv \lim_{n\rightarrow \infty} \sum_{s=0}^{\infty}(-1)^{n+s}\frac{n!\Gamma(2\nu+n+1)}{s!\Gamma(-\nu-n-s+1)}\Delta^{-\nu-n}b_{-n-2s}.
\end{equation}
Notably, all the above equations remain unchanged from Refs.~\cite{BoGao2004,PhysRevA.80.012702} even when $\nu$ is a real non-quater integer value ($s_\ell^2 >0$), or a purely imaginary value  ($s_\ell^2 <0$). 

For the van der Waals QDT expansion, or more generally for real $\nu$, $\nu_0$, and $s_\ell$ values, \equref{M_low_energy} suggests 
\begin{equation}
\label{eq:M_vdw_QDTexp} M_{\ell} \propto |\Delta|^{2\nu_0}\rightarrow 0.
\end{equation}
Therefore, $M_{\ell} $ becomes small at low energy, and thus can be negligible in evaluating the right-hand side of \equref{chi_func} at the leading order~\cite{PhysRevA.62.050702}
 \begin{equation}
\label{eq:threshold_higherZeroE} K^c\simeq \tan\dfrac{\pi\nu_0}{2}.
 \end{equation}

  On the other hand, in the presence of the Efimov effect ($\nu_0 =s_\ell/ 2 $ with $s_\ell^2<0$), $\nu$ takes a purely imaginary value. Using 
\begin{equation}
   \left|\Gamma(1+i|s_\ell|)\right|^2 = \frac{\pi |s_\ell|}{\sinh\pi|s_\ell|}         
\end{equation}
and introducing
\begin{equation}
\label{eq:xil_def}  \xi_\ell\equiv \arg\left[\Gamma\left(\frac{i|s_\ell|}{2}\right)\Gamma(1+i|s_\ell|)\right], 
\end{equation}
we now find 
\begin{equation}
\label{eq:M_Efimov_QDTexp}    M_{\ell}\simeq -|\Delta|^{i|s_\ell|}e^{-2i\xi_\ell}=-\exp\left(i\left[|s_\ell| \ln|\Delta| - 2\xi_\ell\right] \right).
\end{equation}
In stark contrast to \equref{M_vdw_QDTexp}, the magnitude of $M_\ell$ remains unity at low energy. The log-periodic oscillation is a manifestation of the log-periodicity of the Efimov effect; indeed, substituting Eqs.~(\ref{eq:M_Efimov_QDTexp}) (\ref{eq:nu_QDT_expansion}) (\ref{eq:theta_QDT_expansion}) into \equref{chi_func}, we find
\begin{equation}
   \label{eq:chi_func_appendix1}
   K^c \simeq \tan\left[\frac{|s_\ell|}{2}\ln|\Delta|-\xi_\ell\right]\tanh\frac{\pi |s_\ell|}{4} .
\end{equation}
From above, we obtain the binding energy of the Efimov states as a function of $K^c$
 \begin{equation}
 \label{eq:BE_Kc_appendix}
  |\Delta|= \exp\left[\frac{2}{|s_\ell|}\left\{\arctan\left(\frac{K^c}{\tanh\dfrac{\pi|s_\ell|}{4}}\right)+\xi_\ell \right\}\right]e^{-\frac{2n\pi}{|s_\ell|}},
  \end{equation}
which reproduces the discrete scale invariant energy spectrum. The above equation is valid at low energy $|\Delta| \ll 1$, which implies that it holds well when the integer $n$ is large $n \gg 1$.

\section{\label{app:perturbation}Perturbative calculation at weak dipole limit $|a_{dd}|\ll r_{\mathrm{vdw}}$}
When the dipole-dipole interaction is small, we can regard the dipole interaction as a small perturbation for the van der Waals Efimov states in \secref{qdt_vdw}. We start from
\begin{equation}
   \left[\hat{H}_0+\hat{V}_{dd}\right]\ket{\psi_{IL}}=E_{IL}\ket{\psi_{IL}}.
\end{equation}
As the angular momentum is the good quantum number when $\hat{V}_{dd}=0$, we label the states with $L$ and $I$ as $\ket{\psi_{IL}}\simeq \ket{\psi^{(0)}_{IL}}$ : $L$ is the angular momentum of the unperturbed state and $I$ is the label of the state. The second-order perturbation theory reads
\begin{align}
   E_{IL} & =E^{(0)}_{IL}+E^{(1)}_{IL}+E^{(2)}_{IL},         \\
   \ket{\psi_{I L }}      & =\ket{\psi^{(0)}_{IL}}+\sum_{i\ell \neq IL}a^{(1)}_{IL,i\ell}\ket{\psi^{(0)}_{i\ell}}+\sum_{i\ell \neq IL}a^{(2)}_{IL,i\ell}\ket{\psi^{(0)}_{i\ell}},
\end{align}
where
\begin{align}
   E^{(1)}_{IL} & =\bra{\psi^{(0)}_{IL}}\hat{V}_{dd}\ket{\psi^{(0)}_{IL}},    \\
   E^{(2)}_{IL} & =-\sum_{i\ell\neq IL}\frac{\left|\bra{\psi^{(0)}_{IL}}\hat{V}_{dd}\ket{\psi^{(0)}_{i\ell}}\right|^2}{E^{(0)}_{i\ell}-E^{(0)}_{IL}}, \\
   a^{(1)}_{{IL},i\ell}             & =-\frac{\bra{\psi^{(0)}_{i\ell}}\hat{V}_{dd}\ket{\psi^{(0)}_{IL}}}{E^{(0)}_{i\ell}-E^{(0)}_{IL}},       \\
   \begin{split}
      a^{(2)}_{{IL},i\ell}&=-\frac{\bra{\psi^{(0)}_{i\ell}}\hat{V}_{dd}\ket{\psi^{(0)}_{IL}}\bra{\psi^{(0)}_{IL}}\hat{V}_{dd}\ket{\psi^{(0)}_{IL}}}{\left(E^{(0)}_{i\ell}-E^{(0)}_{IL}\right)^2} \\
      &+\sum_{i'\ell' \neq IL}\frac{\bra{\psi^{(0)}_{i\ell}}\hat{V}_{dd}\ket{\psi^{(0)}_{i'\ell'}}\bra{\psi^{(0)}_{i'\ell'}}\hat{V}_{dd}\ket{\psi^{(0)}_{IL}}}{\left(E^{(0)}_{i\ell}-E^{(0)}_{IL}\right)\left(E^{(0)}_{i'\ell'}-E^{(0)}_{IL}\right)} .
   \end{split}
\end{align}
The unperturbed state is $\ket{\psi^{(0)}_{i\ell}} = \dfrac{u_{i \ell}(r)}{r} Y_{\ell m}(\theta,\phi)$, where $u_{i \ell}$ is the solution of the unperturbed Hamiltonian (see \equref{SingleChanBOeq_vdwonly})
\begin{equation}
\left[-\frac{1}{M} \frac{d^2}{dr^2}+ \frac{s_\ell^2-\frac{1}{4}}{Mr^2}-\frac{C_6}{r^6}\right]u_{i\ell}(r)=E_{i \ell}^{(0)}u_{i\ell}(r)
\end{equation}
satisfying the ortho-normalization condition $\displaystyle \int dr u_{i \ell}^*(r) u_{i' \ell}(r) = \delta_{i i'}$. As explained in \secref{qdt_vdw} and Appendix~\ref{app:vdw2body}, $u_{i\ell}$ can be obtained with the QDT.

The matrix element can be evaluated as
\begin{equation}
   \begin{split}
      &\bra{\psi^{(0)}_{i\ell}}\hat{V}_{dd}\ket{\psi^{(0)}_{IL}} \\
      &=\frac{3a_{dd}}{M}\int dr \frac{u^*_{i\ell}(r) u_{IL}(r)}{r^3} \int d\Omega Y_\ell^{m*} (1-3\cos^2\theta) Y_L^m \\
      &=\frac{3a_{dd}}{M}\int_{R_{\mathrm{min}}}^\infty dr \frac{u^*_{i\ell}(r) u_{IL}(r)}{r^3} \\
       & \times2(-1)^{m+1}\sqrt{(2L+1)(2\ell+1)}\left(\begin{array}{ccc}
            \ell & 2 & L \\
            -m   & 0 & m
         \end{array}\right)
      \left(\begin{array}{ccc}
            \ell & 2 & L \\
            0    & 0 & 0
         \end{array}\right),
   \end{split}
\end{equation}
where the bracket in the last line is the Wigner's $3j$ symbol. We note that the azimuthal number $m$ around the $z$ axis is conserved even in the presence of the dipole interaction and thus unchanged. As the $3j$ symbol only takes non-zero values for $\ell=L,L\pm2$, the perturbation couples only $+2, 0, -2$ angular momentum states. Introducing the matrix element of the radial part
\begin{equation}
\label{eq:app_def_udmatel}\langle i,\ell |v_r| I,L \rangle =\int_{R_{\mathrm{min}}}^\infty dr \frac{u^*_{i\ell}(r) u_{IL}(r)}{r^3} 
\end{equation}
for the $L=0$ state corresponding to bosonic heavy atoms, we obtain
\begin{small}
\begin{align}
   \begin{split}
     \label{eq:DeltaE_appEq_boson} E_{I,L=0}&=E^{(0)}_{I,L=0}-\frac{4}{5} \left(\frac{3 a_{dd}}{M}\right)^2\sum_{i}\frac{\left| \langle i,\ell=2|v_r|I,L=0 \rangle \right|^2}{E_{i,\ell=2}^{(0)}-E_{I,L=0}^{(0)}},
   \end{split}                                        \\
   \label{eq:DeltaPsi_appEq_boson}\ket{\psi_{I,L=0}} & =\ket{\psi^{(0)}_{I,L=0}}+\dfrac{2}{\sqrt{5}}\frac{3 a_{dd}}{M} \displaystyle \sum_{i}\dfrac{\langle i,\ell=2|v_r|I,L=0 \rangle }{E_{i,\ell=2}^{(0)}-E_{I,L=0}^{(0)}}\ket{\psi^{(0)}_{i,\ell=2}},
\end{align}
\end{small}
while for the $L=1$ state corresponding to fermionic heavy atoms, we obtain
\begin{small}
\begin{align}
   \label{eq:DeltaE_appEq_fermion}E_{I,L=1}  =& E^{(0)}_{I,L=1}- \frac{4}{5}\frac{3 a_{dd}}{M} \langle I,L=1|v_r| I,L=1\rangle, \\
   \begin{split}
      \label{eq:DeltaPsi_appEq_fermion}\ket{\psi_{I,L=1}}=&\ket{\psi^{(0)}_{I,L=1}}+\frac{3 a_{dd}}{M}\left[\frac{4}{5}\sum_{i\neq I}\frac{ \langle i,\ell=1|v_r|I,L=1\rangle}{E^{(0)}_{i,\ell=1}-E^{(0)}_{I,L=1}}\ket{\psi^{(0)}_{i,\ell=1}}\right. \\
      &\left.+\sqrt{\frac{108}{175}}\sum_{i}\frac{\langle i,\ell=3 |v_r| I,L=1 \rangle }{E^{(0)}_{i,\ell=3}-E^{(0)}_{I,L=1}}\ket{\psi^{(0)}_{i,\ell=3}}\right].
   \end{split}
\end{align}
\end{small}
Notably, the effect of the dipole interaction is qualitatively different for the bosons and fermions: they appear in the second order in $a_{dd}$ for the bosons, and the first order for the fermions. This difference originates from the difference in the matrix element of the dipole interaction: $\langle L M |\hat{V}_{dd}|L M \rangle$=0 for $L=0$ state, while it can take a non-zero value for $L=1$.

\begin{figure}[!t]
	\centering
	\includegraphics[width=1.0\linewidth]{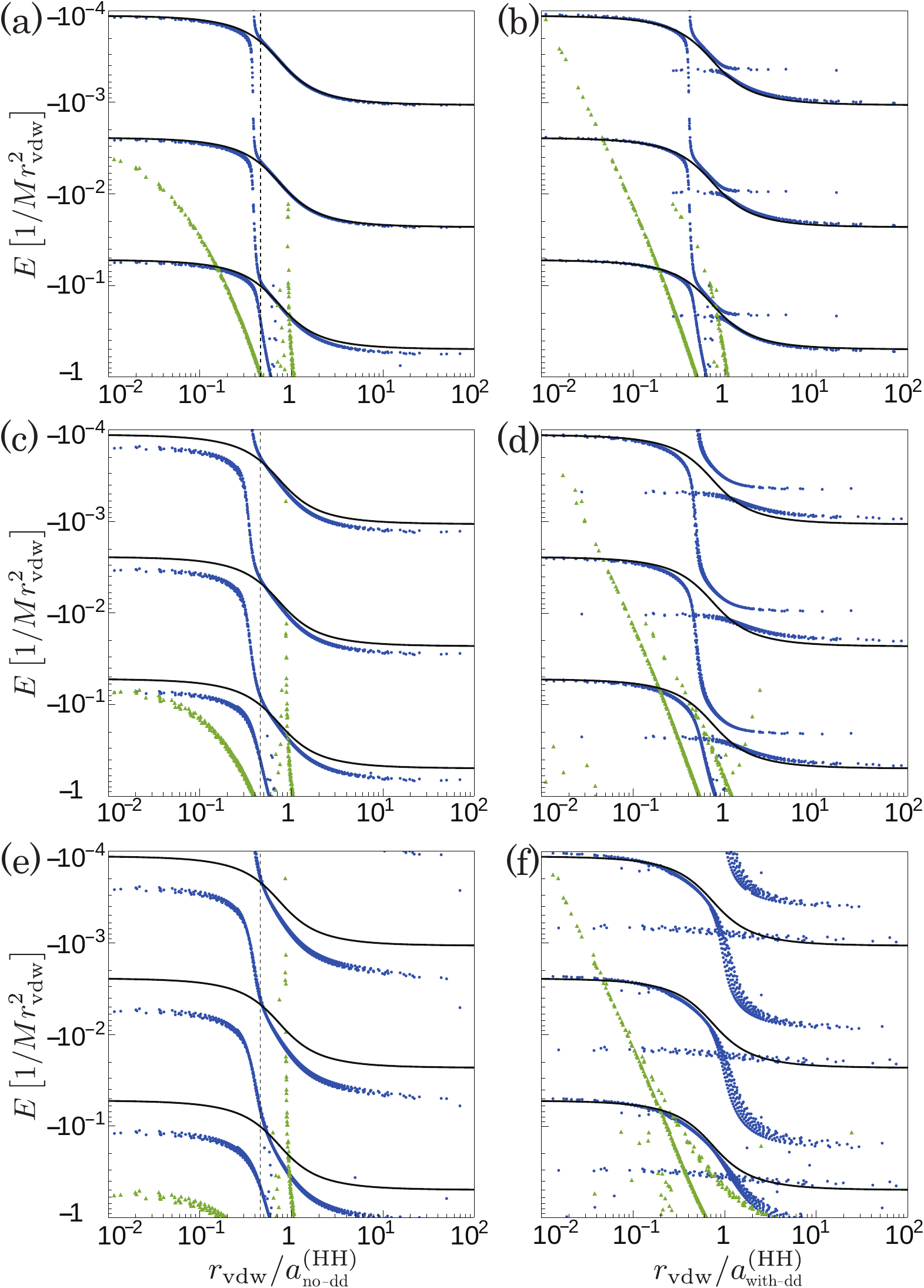} 
\caption{Binding energy of the bosonic dimers of two heavy atoms (green triangles) and trimers (blue circles). The physical parameter in each row is the same as \figref{ModerateDipoleEaHHvHH_Boson}. The left panels are plotted against $a_{\mathrm{no-dd}}$, and the right panels are against $a_{\mathrm{with-dd}}^{(\mathrm{HH})}$, both in their positive scattering length side. The others are the same as \figref{ModerateDipoleEaHHvHH_Boson}.}
\label{fig:AppDimModBoson}
\end{figure}

\section{\label{app:dimer}Dimer energy spectra}
Here, we show the binding energies of the dimers composed of two heavy atoms calculated from \equref{HeavyHeavy2bodyeq}. The numerical parameters and grid points are taken to be the same values as those adopted in \secref{Results}.

For a moderate dipole interaction strength $a_{dd}\sim r_{\mathrm{vdw}}$. the dimer energies for the bosonic system are shown as green points in \figref{AppDimModBoson}. While the dimer energies plotted against $a_{\mathrm{no-dd}}^{(\mathrm{HH})}$ seem to differ significantly from those in the pure van der Waals system (green dashed curve in \figref{VdWOnlyEaHHvHH}(b), they can be converted into a more universal form by introducing the $s$-wave scattering length incorporating the channel couplings $a_{\mathrm{with-dd}}^{(\mathrm{HH})}$, in the same manner as \secref{res_moderate}. This universal behavior of the two-body dipolar scattering is consistent with Refs.~\cite{PhysRevA.64.022717,bohn2009quasi}: the energy of the dimer of $s$-wave dominant character is universally described by the $s$-wave scattering length $E \propto a^{-2}$. There also appear dimers of higher angular momentum, adjacent to the $s$-wave one; they behave universally with respect to different values of $R_{\mathrm{min}}$, but not as much as compared to the $s$-wave one. We have found that the spreading of the data points decrease as $R_{\mathrm{min}}$ is decreased, suggesting that this higher-angular-momentum dimer gradually converges to a universal behavior, with a slowness of convergence similar to that of the trimers in the avoided crossing region. We also show the trimer energies (the same data points in \figref{ModerateDipoleEaHHvHH_Boson}) for comparison. As in \figref{VdWOnlyEaHHvHH}(b), some of the trimers' energies lie above those of the dimers, so that the trimers are embedded in the dimer+atom continuum. We note that the trimers appear as bound states in our Born-Oppenheimer calculation because the continuum channels are neglected. There are two notable dimer curves; one of them is the $s$-wave dominant dimer, which behaves in the same manner as that in \figref{AppDimModBoson} when plotted via $a_{\mathrm{with-dd}}^{(\mathrm{HH})}$. The other one is of  $d$-wave dominant nature, which appears at around  $a_{\mathrm{no-dd}}^{(\mathrm{HH})}/ r_{\mathrm{vdw}} \simeq 1$, and seems to lie more or less besides the trimers in the avoided crossing region if plotted via $a_{\mathrm{with-dd}}^{(\mathrm{HH})}$.

\begin{figure}[!t]
	\centering
	\includegraphics[width=1.0\linewidth]{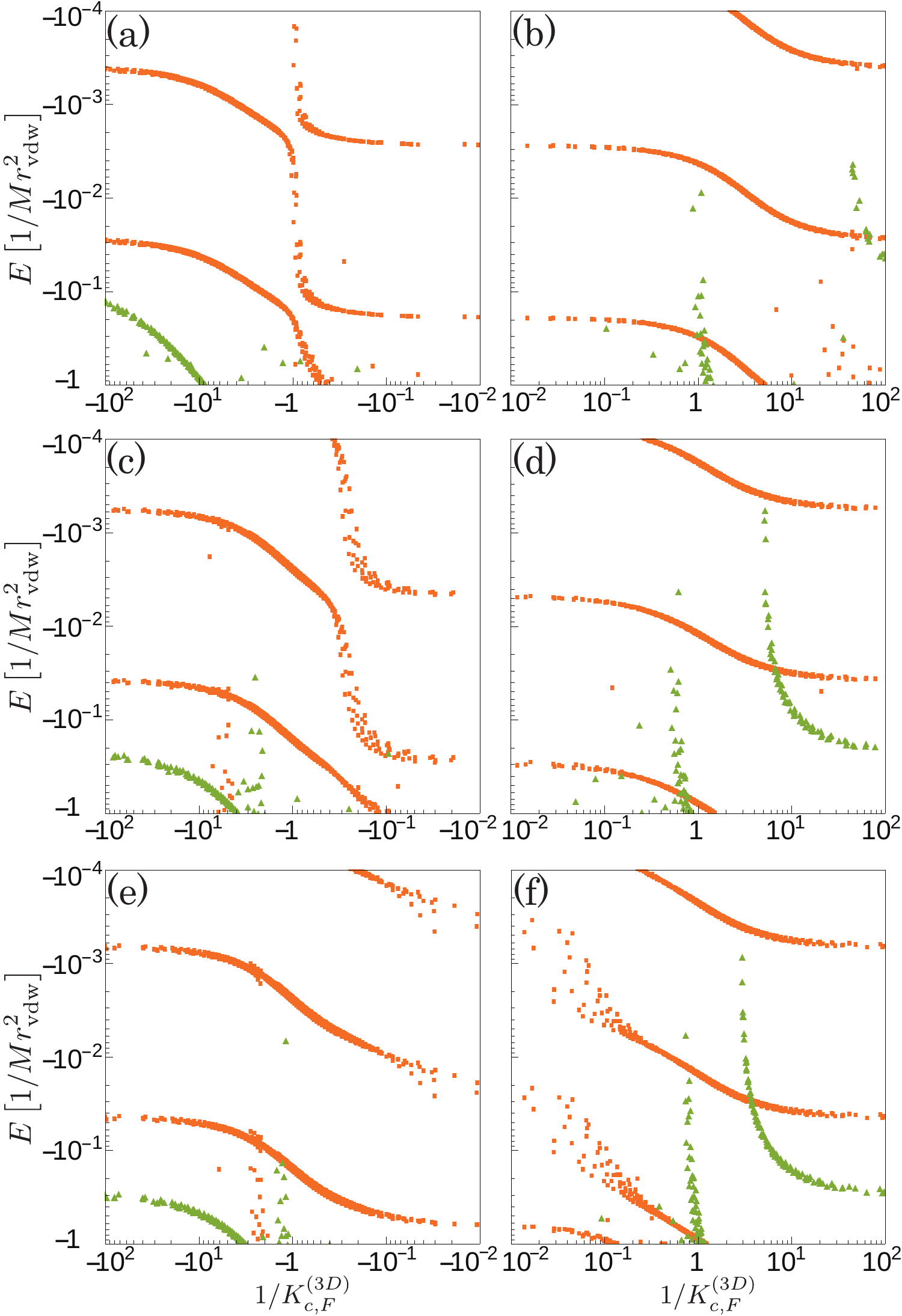} 
\caption{Binding energy of the fermionic dimers of two heavy atoms (green triangles) and trimers (orange squares). The physical parameter in each row is the same as \figref{ModerateDipoleEvHHK_Fermi}.}
\label{fig:AppDimModFermion}
\end{figure}

\begin{figure}[!t]
	\centering
	\includegraphics[width=1.0\linewidth]{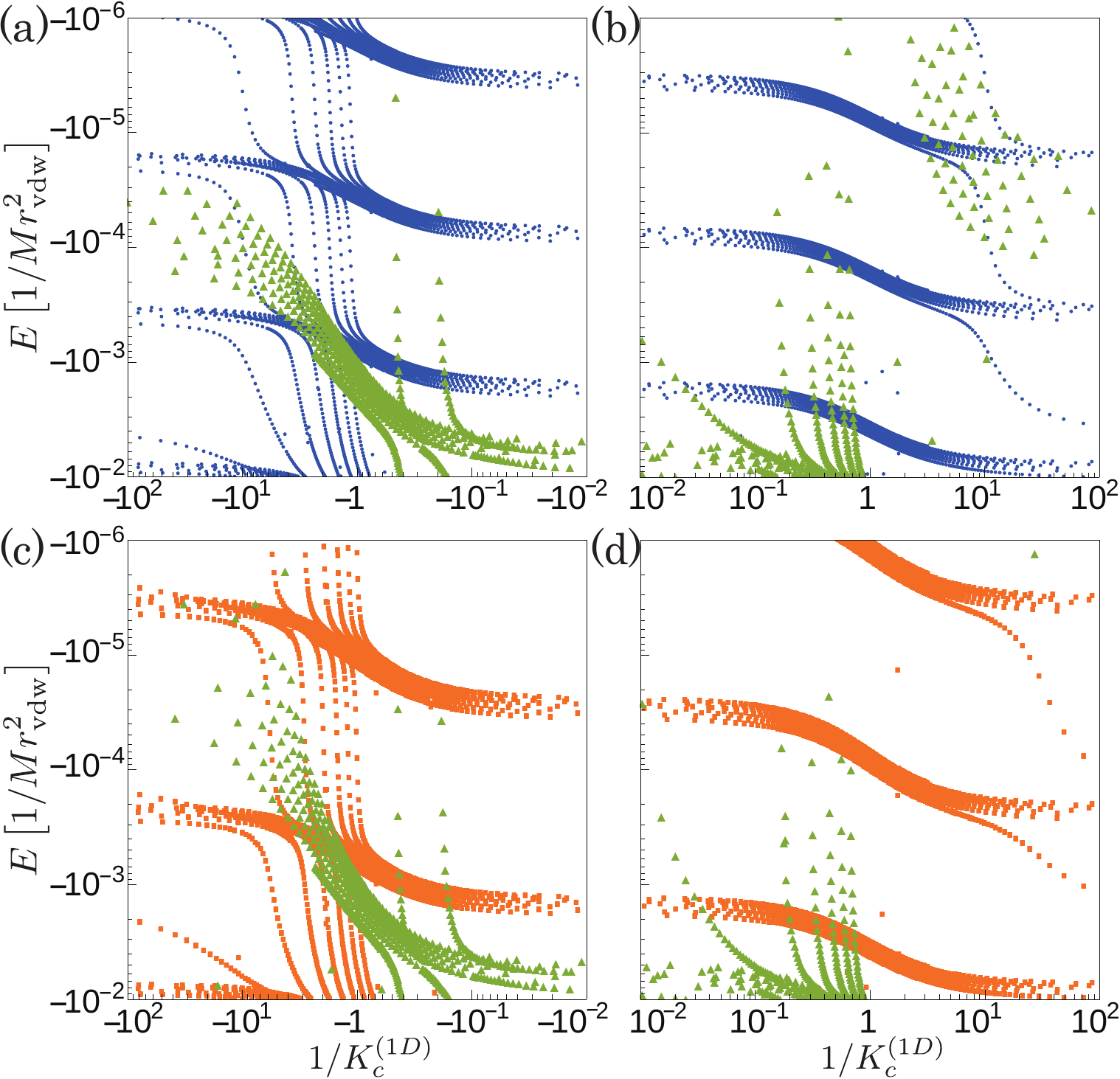} 
\caption{Binding energy of the dimers (green triangles) of two heavy atoms for strong dipole interaction $a_{dd}/r_{\mathrm{vdw}}=50$ (corresponding to the top row of \figref{StrongDipoleEK}). (a)(b) Bosonic system's dimer (green triangles) and trimers (blue circles) in the even parity state, and (c)(d) fermionic system's dimer (green triangles) and trimer (orange squares) in odd parity state. The trimer energies are the same as those in \figref{StrongDipoleEK}(a)(b).}
\label{fig:AppDimStrong}
\end{figure}

In \figref{AppDimModFermion}. dimer energies for the fermionic system are shown, for the parameters corresponding to \figref{ModerateDipoleEvHHK_Fermi}. As it is impossible to introduce the $p$-wave scattering volume, it is not easy to present it in an analogous way to \figref{VdWOnlyEaHHvHH}(d). Still, most of the dimer energies show universal behaviors with respect to $K_{c,F}^{(3D)}$, together with some other stray points which we suspect to be of high-angular-momentum character $\ell \ge 3$. Similarly to \figref{VdWOnlyEaHHvHH}(d), some of the trimers lie above the dimer energy, and are thus embedded in the dimer-atom continuum.

The dimer energies for the strong dipole interaction strength $a_{dd}\gg r_{\mathrm{vdw}}$ are shown in \figref{AppDimStrong}. The parameters are taken to be the same as those in \figref{StrongDipoleEK}(a)(b), and the trimer data are also presented for comparison. While the overall features become more complicated than the moderate one in Figs.~\ref{fig:AppDimModBoson} and~\ref{fig:AppDimModFermion}, possibly due to strong admixture of various angular momentum channels, the overall relative features between the dimers and trimers share similar features.

\clearpage

%

\end{document}